\documentclass[reqno,a4paper]{amsart}
\pdfoutput=1 
\usepackage{amssymb,latexsym} %loads full set of maths symbols.
\usepackage[foot]{amsaddr} %Puts address on title page - foot option puts it as a footnote.
\usepackage{xfrac}
\usepackage{graphicx}
\usepackage[numbers,sort]{natbib}
\RequirePackage{doi} %Converts doi to clickable link.
\usepackage{hyperref} %Enables pdf toc, etc. Use backref option to see where references are used.
\usepackage{xcolor}
\theoremstyle{remark}
\newtheorem{remark}{Remark}[section]

\begin{document}
\title[One  step  RSB and overlaps between two temperatures]{One  step  replica symmetry breaking and overlaps between two temperatures}

\author{Bernard Derrida$^{1,2}$}
\address{$^1$Coll\`ege de France\\ 11 place Marcelin Berthelot\\ 75005 Paris\\ France}
\address{$^2$Laboratoire de Physique de l'Ecole Normale Sup\'erieure\\ ENS\\ Universit\'e PSL\\ CNRS, Sorbonne Universit\'e\\ Universit\'e de Paris\\ F-75005 Paris\\ France}
\email{bernard.derrida@college-de-france.fr}

\author{Peter Mottishaw$^3$}
\email{peter.mottishaw@ed.ac.uk}
\address{$^3$SUPA\\ School of Physics and Astronomy\\ University of Edinburgh\\
	Peter Guthrie Tait Road\\ Edinburgh EH9 3FD\\ United Kingdom}

\begin{abstract}

We obtain an exact analytic expression for the average distribution, in the thermodynamic limit, of overlaps between two copies of the same random energy model (REM) at different temperatures. We quantify the non-self averaging effects and provide an exact approach to the computation of the fluctuations in the distribution of overlaps in the thermodynamic limit. We show that the overlap probabilities satisfy recurrence relations that generalise Ghirlanda-Guerra identities to two temperatures.

We also analyse the two temperature REM using the replica method. The replica expressions for the overlap probabilities satisfy the same recurrence relations as the exact form. We show how a generalisation of Parisi's replica symmetry breaking ansatz is consistent with our replica expressions. A crucial aspect to this generalisation is that we must allow for fluctuations in the replica block sizes even in the thermodynamic limit. This contrasts with the single temperature case where the extremal condition leads to a fixed block size in the thermodynamic limit. Finally, we analyse the fluctuations of the block sizes in our generalised Parisi ansatz and show that in general they may have a negative variance.
\end{abstract}

\maketitle

\section{Introduction}

 Since  replica symmetry  breaking (RSB) was invented by Parisi, 40 years ago \cite{Parisi_1979_Infinite}, it has been used in  many different contexts and the subtle physical meaning of the  scheme he used  has been elucidated         \cite{Parisi_1983_Order,Mezard_1984_Nature,Mezard_1984_Replica}
 (for reviews see \cite{Mezard_1987_Spin} or   \cite{Mezard_2009_Information}). Here we would like to provide a simple example to explore how the Parisi scheme could  be extended to calculate correlations between different temperatures.
	
In the replica approach, a central role is played by the overlaps which represent the
 correlations between pure states. For  a system of $ N $  Ising spins with  the  interactions sampled from some disorder distribution (as in the Sherrington-Kirkpatrick model \cite{Sherrington_1975_Solvable}, for example),  the overlap between a configuration $ \mathcal{C} $ and a configuration $ \mathcal{C}' $  is defined by
\begin{equation} \label{eq:qcc-dfnn}
	q(\mathcal{C},\mathcal{C}') = \frac{1}{N} \sum_{i=1}^{N} \sigma_i^\mathcal{C} \sigma_i^{\mathcal{C}'}
\end{equation}
where $ \sigma_i^\mathcal{C} = \pm 1 $ is the value of the spin at site $ i $ in configuration $ \mathcal{C} $. The  distribution $P(q)$ of this overlap at a single inverse temperature $ \beta $ for a particular sample is then given by \cite{Parisi_1983_Order}
\begin{equation} \label{p-single-temp-dfnn}
	P(q)= \sum_{\mathcal{C},\mathcal{C}'} 
		\frac{e^{-\beta E(\mathcal{C})}}{Z(\beta)} 
		\frac{e^{-\beta E({\mathcal{C}'})}}{Z(\beta)}
		\delta \left(q- q(\mathcal{C},\mathcal{C}')\right) ,
\end{equation} 
where $ Z(\beta) = \sum_{\mathcal{C}}e^{-\beta E(\mathcal{C})}$ is the partition function at inverse temperature $ \beta $ and $ E(\mathcal{C}) $ is the energy of configuration $ \mathcal{C} $ for the particular sample. In a disordered system the energies are quenched random variables and $P(q)$ is itself a random quantity, sample dependent in the sense that it depends on the energies $E(\mathcal{C})$. One of the achievements  of Parisi's theory of spin glasses was to predict   that $P(q)$ remains  sample dependent even in the thermodynamic limit,  and to allow the calculation of various averages and moments which characterize its  sample to sample  fluctuations \cite{Mezard_1984_Nature,Mezard_1984_Replica,Mezard_1985_Random,Derrida_1985_Sample}.

The notion of  overlap distribution  can be generalized when the two configurations are  at different  temperatures 
\begin{equation} \label{p-beta-betaprime-dfnn}
	P_{\beta,\beta'}(q)= \sum_{\mathcal{C},\mathcal{C}'} 
		\frac{e^{-\beta E(\mathcal{C})}}{Z(\beta)}
		\frac{e^{-\beta' E({\mathcal{C}'})}}{Z(\beta')}
		\delta \left(q- q(\mathcal{C},\mathcal{C}')\right) . 
\end{equation}
This clearly reduces to (\ref{p-single-temp-dfnn}) when  $ \beta=\beta' $. These multiple temperature overlaps have mostly been studied in the context of temperature chaos, in order to see how the random free energy  landscapes are correlated  at different temperatures (see for example \cite{Rizzo_2009_Chaos}). Several spin glass models  exhibit temperature chaos, meaning that  the overlap between different temperatures vanishes in the thermodynamic limit  (in which case, the question  of the fluctuations of $P_{\beta,\beta'}(q)$ becomes superfluous). One way to predict temperature chaos is to show that $ P_{\beta,\beta'}(q) $ vanishes  exponentially with the system size when $ \beta \ne \beta' $ and $q>0$ \cite{Franz_1995_Recipes}. There are  however models for which  these multiple temperature overlaps do not vanish and the question of how the Parisi theory has to be modified in these cases is, to our knowledge, not fully understood (see for example the multi-p-spin models analysed in \cite{Rizzo_2002_Ultrametricity}). Here we attack this question in the simplest model which exhibits RSB, the random energy model (REM, see \cite{Derrida_1980_Random,Derrida_1981_Random}) which has the advantage of being open to both exact and replica analysis. This will allow us to propose a way to adapt Parisi's scheme for the two temperature case, in order to be  compatible with our   exact results of section 2. The absence of chaos in the REM has been discussed in \cite{Franz_1995_chaos} as well as its dynamical effects such as rejuvenation in \cite{Sales_2001_Rejuvenation}. 

One of our motivations is the hope that the insight developed here into the use of the replica method in two temperature situations will be applicable to spin glass problems where there is no alternative to the replica method such as the multi-p-spin spherical models discussed in \cite{Rizzo_2002_Ultrametricity}.

In the REM, the energies $E(\mathcal{C})$ are $2^N$ independent random variables distributed according to a Gaussian distribution of width proportional to $N$. The overlap can then only take two values
\[
	q(\mathcal{C},\mathcal{C}')=\delta_{\mathcal{C},\mathcal{C}'} \ .
\]
Therefore  $ P(q) $ consists of two delta function peaks \cite{Gross_1984_simplest,Mezard_1987_Spin},
\begin{equation} \label{eq:Pq-single-T}
	P(q)= (1-Y_{2}) \,\delta \left(q\right) + Y_{2} \, \delta \left(q- 1\right)
\end{equation}
where $Y_{2}$ is the probability, at equilibrium, of finding two copies of the same sample in the same
configuration.
\begin{equation}
	Y_{2}=\sum_{\mathcal{C}}
		\left(\frac{e^{-\beta E(\mathcal{C})}}{\sum_{\mathcal{C}}e^{-\beta E(\mathcal{C})}}\right)^{2}
		=\frac{Z(2\beta)}{Z(\beta)^{2}}\label{P_2_def}.
\end{equation}	
In the large $N$ limit, $Y_{2}$ vanishes in the high temperature phase ($\beta < \beta_c$), while in the low temperature phase ($\beta > \beta_c$) it takes non zero values with sample to sample fluctuations.

A direct calculation \cite{Mezard_1985_Random,Derrida_1985_Sample} as well as a replica calculation \cite{Mezard_1984_Replica} lead to
\begin{equation} \label{Y2} 
	\langle Y_2 \rangle = 1 - \mu  \ \ \ \ \ ; \ \ \ \ \ 
		\langle Y_2^2\rangle - \langle Y_2 \rangle^2 = {\mu-\mu^2 \over 3}  = { \langle Y_2 \rangle - \langle Y_2\rangle^2 \over 3}
\end{equation}
where $\langle . \rangle$  denotes the disorder  average i.e. the average over the random energies $E({\mathcal{C}}) $  and 
\begin{equation} \label{mu-def}
	\mu = {\beta_c \over \beta}.
\end{equation}

The quantity $Y_{2}$ can be generalized to the probabilities
$Y_{k}$ of finding $k$ copies of the same sample in the same configuration
\begin{equation} \label{eq:Yk-definition}
	Y_{k}=\sum_{\mathcal{C}}
		\left(\frac{e^{-\beta E(\mathcal{C})}}{\sum_{\mathcal{C}}e^{-\beta E(\mathcal{C})}}\right)^{k}
		=\frac{Z(k\beta)}{Z(\beta)^{k}} \ . 
\end{equation}
As for $Y_{2}$, the large $N$ limits of the disorder averages of these overlaps
are known \cite{Mezard_1984_Nature,Mezard_1984_Replica,Ruelle_1987_Mathematical,Derrida_1997_random}
\begin{equation} \label{eq:yk-1-temp}
	\langle Y_{k}\rangle=\frac{\Gamma(k-\mu)}{\Gamma(1-\mu)\ \Gamma(k)} \ . 
\end{equation}
Since $\mu= 1 - \langle Y_2 \rangle$, equation (\ref{eq:yk-1-temp}) implies 
\begin{equation} \label{eq:Yk-recursion}
(k \beta - \beta_c  ) \,\left< Y_{k} \right> 
	= k \beta \left< Y_{k+1} \right>
\end{equation}
which  can be seen as  simple cases of the Ghirlanda-Guerra identities \cite{Ghirlanda_1998_General,Bovier_2003_Rigorous,Bovier_2006_Statistical,Panchenko_2016_Chaos}. 

At two different temperatures, the  overlap distribution \eqref{p-beta-betaprime-dfnn} for the REM is still a sum of two delta functions
\begin{equation} \label{eq:Pq-beta-beta}
	P_{\beta,\beta'}(q)= (1-Y_{1,1}) \,\delta \left(q\right) + Y_{1,1} \, \delta \left(q- 1\right) 
\end{equation}
where the random variable 
\begin{equation} \label{eq:y-kk-dfnn}
Y_{k,k'} = \sum_{\mathcal{C}}
        \left(\frac{e^{-\beta E(\mathcal{C})}}{Z(\beta)}\right)^k
        \left(\frac{e^{-\beta' E(\mathcal{C})}}{Z(\beta')}\right)^{k'}
       	= \frac{Z(k \beta + k' \beta')}{Z(\beta)^k \, Z(\beta')^{k'}}
\end{equation}
is the probability that $ k $ copies of a given sample at temperature $ \beta $ and $ k' $ at temperature $ \beta' $ are all in the same configuration.

In section \ref{sec:direct-calculation} and in the Appendix \ref{sec:appendix-direct calculation} we will derive the following exact expressions of the sample averages of these generalized overlaps
\begin{equation} \label{F7a}
	\langle Y_{k,k'} \rangle
	= {\beta \over \beta_c} \ {1   \over \Gamma(k)}\  {1 \over \Gamma(k')}
	\int_0^\infty dv  \ v^{k'-1}
	{\Psi\left( k + k' {\beta' \over \beta} ; v  \right)
	\over \big(-\psi(v)\big)}
\end{equation}
where the functions $\psi(v)$ and $\Psi(v)$ are given by
\begin{equation}
\psi(v) = \int_0^\infty  du
 \ ( e^{-u - v u^{\beta' \over \beta}}-1)
\
 u^{-1-{\beta_c \over \beta}}
\label{psi-defi0}
\end{equation}
and
\begin{equation}
\Psi( z \, ; \, v ) = \int_0^\infty  du \
 e^{-u - v u^{\beta' \over \beta}} \
  u^{z-1 -{\beta_c \over \beta}}
\label{Psi-defi0}
\end{equation}

These two functions are generalisations of the Gamma function. Like the Gamma function which satisfies $\Gamma(z+1)= z \Gamma(z)$,  they obey some recursion relations (which can be obtained from \eqref{Psi-defi0} via integrations by parts)  leading to the following relations
\begin{equation}
        (k \beta + k' \beta' - \beta_c  ) \,\left< Y_{k,k'} \right> = k \beta \left< Y_{k+1,k'} \right>
                + k' \beta' \left< Y_{k,k'+1} \right>
\label{F10a}
\end{equation}
which generalize (\ref{eq:Yk-recursion}). 

We will also show that
\begin{equation}
\langle \big(Y_{1,1} \big)^2 \rangle =
{\beta \over \beta_c}
\
 \int_0^\infty dv  \ v
\left[
{\Psi\left( 2 + 2 {\beta' \over \beta} ; v  \right) \over \big(-\psi(v)\big)}
+ {\Psi\left( 1 +  {\beta' \over \beta} ; v  \right)^2 \over \big(-\psi(v)\big)^2}
\right]
\label{variance}
\end{equation}

\begin{figure}[h]
\centerline{\includegraphics[width=12.5cm]{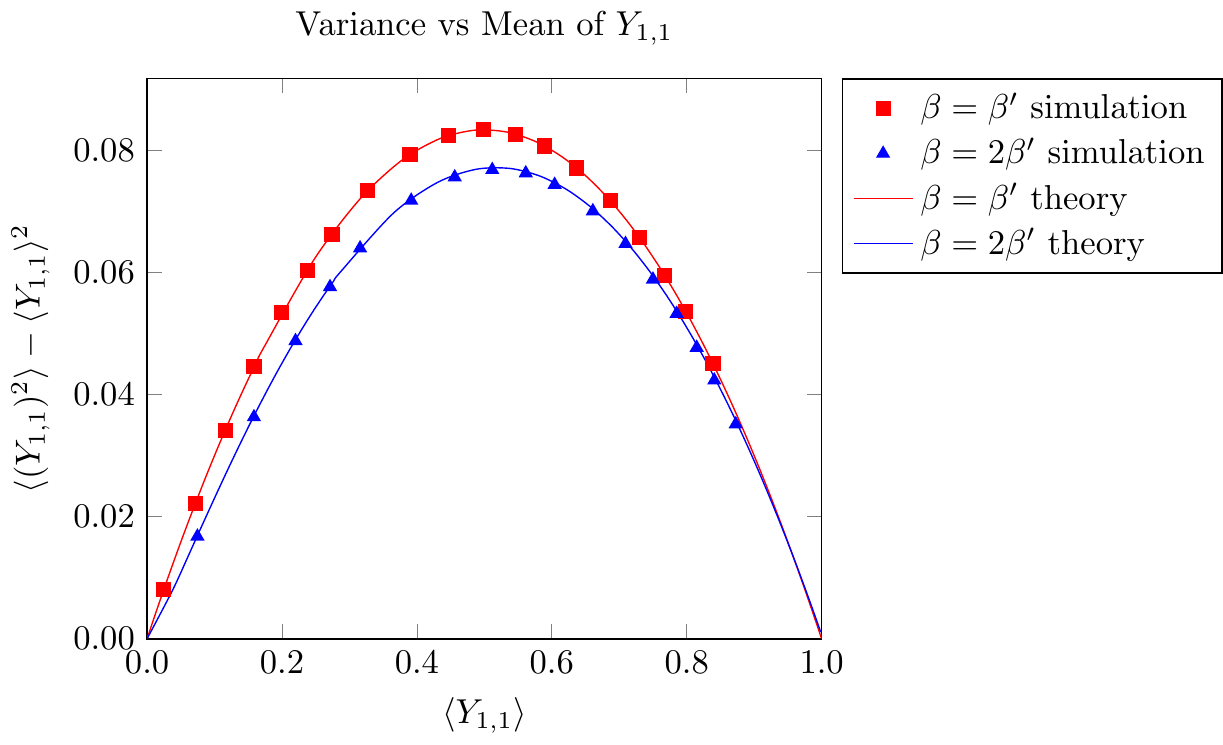}}
\caption{\small  The variance of $Y_{1,1}$ versus its average when  $\beta=\beta'$ and $\beta=2 \beta'$.   The curves are obtained by varying $\beta'$ between $\beta_c$ and $\infty$. The lines represent the expression (\ref{Y2}) in the case $\beta=\beta'$ and (\ref{F7a}) when  $\beta=2 \beta'$. The points are the results of Monte Carlo simulations in these two cases.}
\label{Fig1}
\end{figure}

By varying $\beta$ one can draw, using the exact expressions (\ref{F7a},\ref{variance}),   the variance $\langle \big(Y_{1,1} \big)^2 \rangle - \langle Y_{1,1} \rangle^2$ versus the average $\langle Y_{1,1} \rangle$ as in Figure \ref{Fig1}.  Clearly  the relation (\ref{Y2}) (which is a direct consequence of Parisi's ansatz) is no longer satisfied  when the two temperatures are different ($\beta \neq \beta'$). 

In the rest of the paper we will show in section~\ref{sec:direct-calculation}  how the expressions (\ref{F7a},\ref{psi-defi0},\ref{Psi-defi0},\ref{variance}) can be derived directly. We will also give the generalisation of these expressions when the overlaps are weighted by the partition functions to some power (in the replica language when the number of replicas is non-zero). Then in section~\ref{sec:replica-method}  we will show what needs to be done in order to calculate  these overlaps in a replica approach and in  section~\ref{sec:Parisi-matrices} we will propose a scheme which generalises Parisi's ansatz and is compatible with our exact results. Finally, in section~\ref{sec:Fluctuations} we will explore the nature of the fluctuations in block size that we observe in this generalisation of Parisi's ansatz to two temperatures.

\section{The direct calculation of the overlaps}
\label{sec:direct-calculation}
In this section, after recalling the definition of the REM, we explain how the expressions of the overlaps such as  (\ref{F7a},\ref{psi-defi0},\ref{Psi-defi0})
can be derived.
In the REM, a sample is determined by the choice of $2^N$ random energies $E(\mathcal{C})$ chosen independently from a Gaussian distribution
\begin{equation} \label{eq:rem-disorder}
	P(E) = {1 \over \sqrt{N \pi J^2 }} \exp\left[ -{E^2 \over N J^2 }\right]
\end{equation}
It is known that,
in the thermodynamic limit ($N \to \infty$),  there is a phase transition
\cite{Derrida_1980_Random,Derrida_1981_Random}
at an inverse temperature
\begin{equation} \label{eq:beta_c}
	\beta_c = \frac{2 \sqrt{\log 2}}{J}
\end{equation}
and that in the frozen phase  $\beta > \beta_c$ (which is the only phase with non-zero overlaps) the partition function is  dominated by energies close to the ground state which itself has fluctuations of order $1$ around a characteristic energy $E_0 = -J N \sqrt{\log 2} + J \log N/(4 \sqrt{\log 2})$.
As only energies at a distance of order $1$  (i.e $\ll N$) contribute to the partition function one can
 replace the REM by a Poisson REM \cite{Derrida_2015_Finite} which  has, in the frozen phase and for large system sizes, the same  properties as the original REM \cite{Derrida_1980_Random}. 
In this Poisson REM (PREM), the values of the energies, for a given sample, are the points generated by a Poisson process \cite{Ruelle_1987_Mathematical} on the real line with intensity 
\begin{equation} \label{poisson}
\rho\left(E\right)= C \exp[\beta_c \, ( E-E_0) ] \ \ \ \  \textrm{with} \ \ \ \ C= {1 \over J \sqrt{\pi}} \ . 
\end{equation}
One way to think of it is to slice the real axis into infinitesimal energy intervals $(E,E+dE)$
indexed by $\nu$,    and to say that there is an energy $E_\nu$ in the interval $\nu$ with probability $p_\nu= \rho(E) dE$. In other words  the partition function is given by
\begin{equation}
Z(\beta)=\sum_{\nu=-\infty }^\infty  y_\nu \exp[-\beta E_\nu]
\label{ZZ}
\end{equation}
where the  $y_\nu$ are independent binary random variables  such that
$y_\nu = 1$ with probability $p_\nu$ and $y_\nu=0$ with probability $1- p_\nu$ (because the intervals are   infinitesimal, there is  no interval $\nu$  occupied by more than one energy).

The details of the calculation leading to the expressions
(\ref{F7a},\ref{psi-defi0},\ref{Psi-defi0}) are given in  Appendix  \ref{sec:appendix-direct calculation}.

One can generalize  (\ref{F7a})  to obtain (see (\ref{F8}) of Appendix \ref{sec:appendix-direct calculation}) the average over disorder of 
$Y_{k_1,k_2; k_1',k_2'}  $ defined by
$$Y_{k_1,k_2; k_1',k_2'} = {\sum_{\nu \neq \nu'} \, y_\nu \,   \ y_{\nu'} \,  e^{
 -(\beta k_1 + \beta' k_1')E_\nu
 -(\beta k_2 + \beta' k_2')E_{\nu'} } \over Z(\beta)^{k_1+k_2} \ Z(\beta')^{k_1'+k_2'}} $$
In particular this allows one to obtain (\ref{variance})  from (\ref{F8}) as one has from the definition (\ref{eq:y-kk-dfnn})
$$(Y_{1,1})^2 = Y_{2,2} + Y_{1,1; 1,1} \ . $$

In the replica approach, as we will see,  it is often  convenient to deal with weighted overlaps
defined as
\begin{equation} \label{eq:ykk-z-moments-ratio}
\langle Y_{k,k'} \rangle_{n,n'} = {\left\langle Z(k \beta + k' \beta')  \ Z(\beta)^{n-k}  \ Z(\beta')^{n'-k'}\right\rangle \over \langle Z(\beta)^{n} \ Z(\beta')^{n'} \rangle }
\end{equation}
(see (\ref{eq:y-kk-dfnn})).
These averages can be   performed (see Appendix \ref{sec:appendix-direct calculation}) to get 
 expressions (\ref{F9},\ref{F10}) (valid for $n<0$ and $n'<0$)  which generalize (\ref{F7a}) and (\ref{F10a})
\begin{equation}
\langle Y_{k, k'}\rangle_{n,n'}
=
  {
(-r)  \ \Gamma(-n) \ \Gamma(-n')
   \over \Gamma(k-n)\   \Gamma(k'-n')}
   { \int_0^\infty dv  \ v^{k'-n' -1}
\Psi\left( k + k' {\beta' \over \beta} ; v  \right)
\ (-\psi(v))^{r-1}
        \over
        \int_0^\infty dv   \ v^{-n'-1}\
        (-\psi(v))^r }
\label{F9a}
\end{equation}
 where 
 \begin{equation} 
r={n \beta + n' \beta' \over \beta_c}  
\label{r-expressiona}
\end{equation}
and
\begin{equation}
	(k \beta  + k'  \beta' - \beta_c ) \,\left< Y_{k,k'} \right>_{n,n'} = (k-n) \beta \left< Y_{k+1,k'} \right>_{n,n'}
		+  (k'-n')\beta' \left< Y_{k,k'+1} \right>_{n,n'}.
\label{F10b}
\end{equation}

\begin{remark}
	{\it The $n,n' \to 0 $ limit}\\*
	As explained in  Appendix  \ref{sec:appendix-direct calculation} (see (\ref{asymptotics})) the expression (\ref{F9a}) reduces to (\ref{F7a}) in the limit $n\to 0^-$ and $n' \to 0^-$. Very much like the integral representation of the Gamma function the expression (\ref{F9a}) would take a different form for $n$ and/or $n' >  0$, and so we would need to use these alternative expressions to verify that  the limits $n \to 0^+$ and $n' \to 0^+$ lead also to (\ref{F7a}).
\end{remark}

\begin{remark}
	{\it The $\beta'=\beta $ case}\\*
	In the particular case where $ \beta=\beta'$, one can perform the integrals in (\ref{psi-defi0},\ref{Psi-defi0}) as in (\ref{equalT1},\ref{equalT2})  and  obtain more explicit expressions of the overlaps as in (\ref{F8a}). In particular one gets  (see (\ref{F8a})) that
	\begin{equation}
	\langle Y_{k,k'} \rangle
	=
	{1  \over \Gamma(k+k')}\    {\Gamma (k+ k'-{\beta_c\over \beta}) \over \Gamma(1-{\beta_c \over \beta}) }
	\label{F8b}
	\end{equation}
	which agrees with  (\ref{eq:yk-1-temp})  as $Y_{k,k'}=Y_{k+k'}$ when $\beta=\beta'$ (see the definitions
	(\ref{eq:yk-1-temp},\ref{eq:y-kk-dfnn})).
	Similarly (\ref{F9a}) becomes when $\beta=\beta'$
	\begin{equation}
	\langle Y_{k,k'} \rangle_{n,n'}
	=
	{\Gamma(1-n-n')  \over \Gamma(k+k'-n-n')}\    {\Gamma (k+ k'-{\beta_c\over \beta}) \over \Gamma(1 -{\beta_c \over \beta}) }
	\label{F9b}
	\end{equation}
	As in this single temperature case $ Y_{k,k'} = Y_{k+k'} $,  \eqref{F10a} reduces to equation~\eqref{eq:Yk-recursion}. Therefore once $\langle Y_2\rangle$ is known, all the other $\langle Y_k \rangle$ can be determined by the relations \eqref{eq:Yk-recursion}. Clearly for $\beta' \neq \beta$ this is not the case.  However, \eqref{F10a} or (\ref{F10b}) would allow to determine all the $\langle Y_{k,k'}\rangle$  from the knowledge of all the $\langle Y_{k,1} \rangle$.
\end{remark}

In the rest of the paper  we will see how  expressions  (\ref{F7a}) or (\ref{F9a}) can be interpreted in terms of  replica symmetry breaking.

\section{The replica method} 	
	\label{sec:replica-method}
In this section we  apply the replica method to the REM. To illustrate the approach we first recall the computation of the free energy and overlap probability $ \big\langle Y_k \big\rangle_n $ for a single temperature. It is well known that a single step in Parisi's replica symmetry breaking scheme \cite{Parisi_1979_Infinite} gives the correct low temperature solution \cite{Derrida_1981_Random,Gross_1984_simplest}. Here we will start with a slightly more general approach that allows for fluctuations in the block sizes. In the single temperature case the need to allow fluctuations in the block sizes has been discussed in \cite{Nieuwenhuizen_1996_Puzzle,Ferrero_1996_Fluctuations} and used in \cite{Campellone_2009_Replica,Derrida_2015_Finite} to compute finite size corrections in the REM. 

In the two temperature case block size fluctuations have also been discussed in the context of temperature chaos in spin glasses (see \cite{Rizzo_2001_chaos} and \cite{Yoshino_2008_Stepwise} Appendix~G for a detailed discussion), but computing the full overlap distribution between two temperatures by averaging over these block fluctuations has proved challenging. In this section we outline a replica symmetry breaking scheme for the two temperature case and show that it satisfies the same recursion \eqref{F10a} as the exact solution. 

\subsection{The REM at a single temperature}

 To implement the replica method, the first step is to calculate the integer moments of the partition function, $ \bigl< Z(\beta)^n \bigr> $, then one assumes that the expression is valid for non-integer $ n $ and finally one makes use of 
\begin{equation} \label{eq:replica-trick}
\bigl< \log Z  \bigr> =  \lim\limits_{n \to 0} \frac{\log \bigl< Z(\beta)^n \bigr>}{n}
\end{equation}
to obtain the disorder average of the free energy. 

\subsubsection{Integer moments of the partition function}
 As shown in Appendix \ref{sec:appendix-integer-moments}  the integer moments of the partition function for the  REM  are given by
\begin{equation} \label{eq:Zn-sum-on-mu-exact}
	\big\langle Z(\beta)^n \big\rangle 
	= \sum_{r \geq 1} \frac{1}{r!} 
	\sum_{\mu_1 \geq 1} \cdots \sum_{\mu_r \geq 1}
	C_{n,r}  \left( \{\mu_i \}\right)
\big\langle Z(\beta \mu_1) \big\rangle \, \big\langle Z(\beta \mu_2) \big\rangle  \cdots \big\langle Z(\beta \mu_r) \big\rangle
% \exp \left( N \left[ A(r, \{\mu_i\})  \right] \right)
\end{equation}
(see (\ref{eq:Zn-sum-on-mu-exact-appendix})
in  appendix~\ref{sec:appendix-integer-moments}) where 
\begin{equation} \label{eq:C-dfnn}
C_{n,r} \left( \{\mu_i \}\right) 
= \frac{n!}{\mu_1! \mu_2! \cdots \mu_r!}
\delta \left[ \sum_{i=1}^{r} \mu_i = n \right]
\end{equation}
and 
 the Kronecker delta 
$ \delta \left[ \sum_{i=1}^{r} \mu_i = n \right] $ 
 ensures that the $ \{\mu_i\}$  sum to $ n $. 
 As 
\begin{equation}
	 \big\langle Z(\beta) \big\rangle
	= \int_{-\infty}^{\infty} \rho(E) e^{-\beta E} dE
= e^{N \, f(\beta)}
\end{equation}
with 
 \begin{equation} \label{eq:f-dfnn}
 	f(\beta) = \log 2 + \frac{(\beta J)^2}{4}.
 \end{equation} 
one can rewrite
(\ref{eq:Zn-sum-on-mu-exact}) as 
\begin{equation} \label{eq:Zn-sum-on-mu-exact-bis}
	\left\langle Z(\beta)^n \right\rangle 
	= \sum_{r \geq 1} \frac{1}{r!} 
	\sum_{\mu_1 \geq 1} \cdots \sum_{\mu_r \geq 1}
	C_{n,r}  \left( \{\mu_i \}\right)
 \ e^{ N  A(r, \{\mu_i\})  }
\end{equation}
where
\begin{equation} \label{eq:A-single-T-dfnn-nu}
	A(r, \{\mu_i\}) = \sum_{i=1}^{r} f(\mu_i \, \beta)
	 = r \log 2 + {(\beta J)^2 \over 4} \sum_{i=1}^{r}  \mu_i^2 \ . 
\end{equation}
One can interpret a given term of  the sum \eqref{eq:Zn-sum-on-mu-exact} or \eqref{eq:Zn-sum-on-mu-exact-bis} as  $ n $ replicas distributed  over $ r $ distinct configurations with $\mu_i$ replicas in configuration $i$. Before using the replica method to compute the free energy from these expressions we compute the overlap probability for integer $ n $.

An expression for the overlap probability $ Y_k $, defined in \eqref{eq:Yk-definition}, can be obtained from the ratio of integer moments
\begin{equation}
\left\langle Y_k \right\rangle_n 
= \frac{\left\langle Y_k \, Z(\beta)^n \right\rangle}
{\left\langle Z(\beta)^n \right\rangle}
= \frac{\left\langle Z(k\beta)\,  Z(\beta)^{n-k} \right\rangle}
{\left\langle Z(\beta)^n \right\rangle}.
\end{equation}
The denominator in the rightmost expression is the single temperature moment in  \eqref{eq:Zn-sum-on-mu-exact-bis}. If $ n $ and $ k $ are positive integers with $ n > k $ then the numerator is a two temperature moment of the partition function that is computed  (see  (\ref{explanation1},\ref{explanation2}) in Appendix ~\ref{sec:appendix-integer-moments}) and one gets
\begin{equation} \label{eq:Yk-mu-average}
\big\langle Y_k \big\rangle_n = 
\left\langle r \frac{\mu_1(\mu_1-1) \cdots (\mu_1-k+1)}
{n(n-1) \cdots (n-k+1)} \right\rangle_{\{\mu_i\}}
\end{equation}
where the average $ \langle . \rangle _{\{\mu_i\}}$ means that  for any  function $ F \left(\{\mu_i\}\right) $, 
\begin{equation}
\left\langle F \left(\{\mu_i\}\right) \right\rangle_{\{\mu_i\}}
= \frac{ \displaystyle{\sum_{r \ge 1} 
		\sum_{\{\mu_i \geq 1\}}
		F \left(\{\mu_i\}\right)
		W_r(\{\mu_i\})}
}{ \displaystyle{\sum_{r \ge 1} 
		\sum_{\{\mu_i \geq 1\}}
		W_r(\{\mu_i\})	
}}.
\end{equation}
with (see (\ref{eq:C-dfnn},\ref{eq:Zn-sum-on-mu-exact-bis}))
$$W_r(\{\mu_i\}) = \frac{C_{n,r} \left( \{\mu_i \}\right)}{r!}
e^{ N  A(r, \{\mu_i\}) }. $$

\subsubsection{The thermodynamic limit and the extremal condition}
\label{1T-thermo}

In the thermodynamic limit $ \left\langle Z(\beta)^n \right\rangle $ in equation~\eqref{eq:Zn-sum-on-mu-exact-bis} should be dominated by terms which maximize  $A(r, \{\mu_i\}) $ in equation~\eqref{eq:A-single-T-dfnn-nu}. At high temperatures the maximum  corresponds to all $ n $ replicas being in different configurations.  Thus  $ r=n $,  $ \mu_i =1 $ for all $ i $. Then
\begin{equation}
	\left\langle Z(\beta)^n \right\rangle \simeq  e^{N  \, n\,  f(\beta)}
\end{equation}
which gives   (see \eqref{eq:replica-trick})
\begin{equation}
\langle \log Z \rangle  = N f(\beta)
	=N \left[\log 2+ \frac{(\beta J) ^2}{4}\right].
\label{logZ}
\end{equation}
The entropy of this solution is
\begin{equation} \label{eq:S-entropy}
	\bigl< S \bigr> = N \left[ f(\beta) - \beta f'(\beta)\right]
	= N \left[ \log 2 -\frac{(\beta J)^2}{4} \right].
\end{equation}
There is a critical inverse temperature $ \beta_c $ where this entropy vanishes. It is the solution of
\begin{equation} \label{eq:beta_c-eqtn}
	f(\beta_c) - \beta_c f'(\beta_c) = 0
\end{equation}
and therefore given by \eqref{eq:beta_c}. When $ \beta < \beta_c $ the entropy in \eqref{eq:S-entropy} is positive and (\ref{logZ}) is indeed the right free energy \cite{Derrida_1980_Random}. On the other hand  at low temperatures, when $ \beta > \beta_c $, the entropy is negative and one  must look for a different solution.

To do so we proceed as Parisi did in his original papers \cite{Parisi_1979_Infinite,Parisi_1980_Sequence,Parisi_1980_order,Parisi_1983_Order} on replica symmetry breaking. To identify the terms that dominate the sum in \eqref{eq:Zn-sum-on-mu-exact-bis} in the thermodynamic limit ($ N \to \infty $) in the low temperature phase we make the following three  assumptions:

\begin{enumerate}
	\item We expect \emph{all} the dominant terms to have a large $ N $ behaviour of the form $ \exp N A(r, \{\mu_i\}) $ with the \emph{same} value of $ A $ and the \emph{same} value of $r$. 
	\item The dominant terms in the $ n \to 0 $ limit correspond to the minimum of $  A(r, \{\mu_i\}) $ and not the maximum. This seems an unreasonable assumption, but gives the correct result when replica symmetry is broken. One  argument to support this assumption is that when the number of independent parameters we are maximising over is negative the maximum becomes a minimum \cite{Mezard_1987_Spin}. In \eqref{eq:A-single-T-dfnn-nu} there are $ r-1 $ independent parameters $ \mu_i $ (due to the constraint $ \sum_{i=1}^{r} \mu_i = n $) and, as we will see in \eqref{eq:r-and-mu-1T} below,  $ r-1 $ is negative when $ n < \frac{\beta_c}{\beta} $.
	\item We  allow $ n, r, \mu_i $ to become real parameters  when we compute the minimum of $  A(r, \{\mu_i\}) $. 
\end{enumerate}
As for Parisi's original ansatz, it is clear that these assumptions, as such, have no rigorous justification. However, the free energy obtained using these assumptions has been verified for a number of spin glass models by a rigorous mathematical analysis (for reviews see \cite{Panchenko_2013_Sherrington,Guerra_2014_Interpolation}).  They also lead to the correct free energy of the REM in the low temperature phase
\cite{Derrida_1981_Random}. In addition, there is a representation of the free energy as a contour integral  in the complex plane \cite{Campellone_2009_Replica,Derrida_2015_Finite}, where the minimum of $ A(r,\{\mu_i\}) $ and the non-integer values of $ r $ and the $ \mu_i $ appear naturally as a consequence of taking the saddle point along the contour.

The minimum of $ A(r, \{\mu_i\})  $ in \eqref{eq:A-single-T-dfnn-nu}
 with respect to the $ \{ \mu_i \} $ subject to the constraint $ \sum_{i=1}^{r} \mu_i =n $ can be found using a Lagrange multiplier and it corresponds to all $ \mu_i $  taking the same value. The constraint then gives immediately
\begin{equation} \label{eq:mu-a-equal-mu}
\mu_i = \frac{n}{r}.
\end{equation}
for all $ i $. Then 
 \eqref{eq:A-single-T-dfnn-nu}  gives $A(r,\{\mu_i\}) = r f({n \beta \over r}) $ and 
taking  the extremal value  
with respect to $ r $  
one gets
\begin{equation}
f\left(\frac{\beta n}{r}\right) 
- \frac{\beta n}{r} f'\left(\frac{\beta n}{r}\right) = 0.
\end{equation}
Comparison with \eqref{eq:beta_c-eqtn} gives 

\begin{equation} \label{eq:r-and-mu-1T}
	r=n \frac{\beta}{\beta_c}, \ \ \ \ \  {\rm so  \ that}  \ \ \ \ \ \mu={\beta_c \over \beta}
\end{equation}
so that for large $ N $ we can approximate \eqref{eq:Zn-sum-on-mu-exact} as
\begin{equation}\label{eq:Zn-large-N}
\bigl< Z(\beta)^n \bigr> 
\sim 
%	\simeq \frac{n!}{ \left(\frac{n}{\mu}\right)! (\mu!)^{\frac{n}{\mu}} }
		\exp \left\{Nn \left[ \frac{1}{\mu} \log2 + \frac{\beta^2}{4} \mu
	\right]\right\} = 
		\exp \left\{Nn \frac{\beta}{2 \beta_c} 
	\right\},
\end{equation}
where we have defined $ \mu = \frac{\beta_c}{\beta} $. The free energy \eqref{eq:replica-trick} in the frozen phase is therefore $ \langle \log Z \rangle =  N {\beta \over 2 \beta_c}  $ which is known to be the correct expression \cite{Derrida_1980_Random}

The extremal condition \eqref{eq:mu-a-equal-mu} tells us that $ \mu_i =\mu $ and $ r= \frac{n}{\mu} $ for the dominant terms. So the $ \mu_i $ do not fluctuate and we can immediately write
\begin{equation} \label{eq:ykn-1-temp}
\left\langle Y_k \right\rangle_n = 
	r \frac{\mu(\mu-1) \cdots (\mu-k+1)}
		{n(n-1) \cdots (n-k+1)}
	= \frac{\Gamma(k-\mu)\, \Gamma(1-n) }{\Gamma(1-\mu) \, \Gamma(k-n)}
\end{equation}
which in the $ n \to 0 $ limit gives the well known result \eqref{eq:yk-1-temp} and confirms that $ \langle Y_2 \rangle = 1- \mu $. We  are now going to see that in the two temperature case this simple last step is not possible because fluctuations  of $\mu_i$ remain even in the thermodynamic limit.

\subsection{The REM at two temperatures}

\subsubsection{Integer moments of the partition function}

In the two temperature case our starting point is the following expression for the moments (see Appendix 
~\ref{sec:appendix-integer-moments})
\begin{multline} \label{eq:ZZnn-sum-on-mu-exact}
\left\langle Z(\beta)^n Z(\beta')^{n'} \right\rangle 
	= \sum_{r \geq 1} \frac{1}{r!} 
	\sum_{\{\mu_i \geq 0\}}
	\sum_{\{\mu'_i \geq 0\}}
	\delta \left[ \mu_i + \mu'_i \geq 1 \right] \\
	\times
	C_{n,r} \left( \{\mu_i \}\right)\,  C_{n',r} \left( \{\mu'_j \}\right)
	\, e^{ N A \left(r, \{\mu_i, \mu'_i \}\right)}
\end{multline}
where the sum on $ \{\mu_i\}, \{\mu'_i\} $ is over all non-negative integers, the  
$ C_{n,r} \left( \{\mu_i \}\right) $ 
and 
$ C_{n',r} \left( \{\mu_i' \}\right) $ 
are  combinatorial factors  defined in \eqref{eq:C-dfnn} and 
\begin{equation} \label{eq:A-two-T-dfn}
A\big(r, \{\mu_i,\mu'_i\}\big) = \sum_{i=1}^{r} f(\mu_i \, \beta +  \mu'_i \, \beta')
 = r \log 2 
	+ \frac{J^2}{4} \sum_{i=1}^{r} \left( \mu_i \,  \beta +  \mu'_i \, \beta'\right)^2
.
\end{equation}
As in (\ref{eq:Zn-sum-on-mu-exact-bis}) each term in the sum (\ref{eq:ZZnn-sum-on-mu-exact}) corresponds to a different grouping of the $n+n'$ replicas: in configuration $i$ there are $\mu_i$ replicas at inverse temperature $\beta$ and $\mu_i'$ replicas at inverse temperature $\beta'$.
There is an additional constraint associated with each configuration $ i $ that $ \mu_i + \mu'_i \ge 1$; in other words we need at least one replica, which can be from either $ n $  or $ n' $, in each configuration.

We are interested in how the single temperature overlap calculation leading to~\eqref{eq:ykn-1-temp} generalises to the two temperature case. We start with the weighted form of  $ Y_{k,k'} $ defined in \eqref{eq:ykk-z-moments-ratio}
\begin{equation}
\left\langle Y_{k,k'} \right\rangle_{n,n'}
= \frac{\left\langle Y_{k,k'} Z(\beta)^n Z(\beta')^{n'} \right\rangle}{\left\langle Z(\beta)^n Z(\beta')^{n'}  \right\rangle}
= \frac{\left\langle Z(\beta k + \beta' k') 
	Z(\beta)^{n-k} Z(\beta')^{n'-k'} \right\rangle}
{\left\langle Z(\beta)^n Z(\beta')^{n'}  \right\rangle}
\end{equation}
The denominator in the rightmost expression is the two temperature moment in \eqref{eq:ZZnn-sum-on-mu-exact}. The numerator is a three temperature moment of the partition function  \eqref{eq:ZZZnnn-sum-on-mu-exact-appendix} computed in appendix~\ref{sec:appendix-integer-moments}.  By a direct generalisation of the derivation (\ref{explanation1},\ref{explanation2})   of (\ref{eq:Yk-mu-average}) one gets
\begin{equation} \label{eq:Ykk-mu-mu-average}
\left< Y_{k,k'} \right>_{n,n'}
= \left< r \;\frac{\mu_1 (\mu_1-1) \cdots (\mu_1 - k +1)}{n(n-1) \cdots (n-k+1)} 
\frac{\mu'_1 (\mu'_1-1) \cdots (\mu'_1 - k' +1)}{n'(n'-1) \cdots (n'-k'+1)}
\right>_{\{\mu_i,\mu'_i\}}
\end{equation}
where the average $ \langle . \rangle _{\{\mu_i,\mu'_i\}}$ means that  for any  function $ F \left(\{\mu_i,\mu'_i\}\right) $, 
\begin{equation} \label{eq:F-mu-mu-average}
	\left\langle F \left(\{\mu_i,\mu'_i\}\right)
		 \right\rangle_{\{\mu_i,\mu'_i\}}
	= \frac{ \displaystyle{\sum_{r \ge 1} 
			\sum_{\{\mu_i \geq 0\}} \sum_{\{\mu'_i \geq 0\}}
			F \left(\{\mu_i,\mu'_i\}\right)
			W_r(\{\mu_i,\mu'_i\})}
	}{ \displaystyle{\sum_{r \ge 1} 
			\sum_{\{\mu_i \geq 0\}} \sum_{\{\mu'_i \geq 0\}}
			W_r(\{\mu_i,\mu'_i\})	
	}}
\end{equation}
with (see (\ref{eq:C-dfnn},\ref{eq:A-two-T-dfn}))
$$W_r(\{\mu_i,\mu'_i\}) = 
	\frac{C_{n,r} \left( \{\mu_i \}\right)
		C_{n',r} \left( \{\mu'_i \}\right)
	}{r!}
	e^{ N  A(r, \{\mu_i,\mu'_i\}) } 
	\prod_{i=1}^{r} \theta \left[ \mu_i + \mu'_i \geq 1 \right].
$$
Here $ \theta \left[ \mu_i + \mu'_i \geq 1 \right] $ is one if $ \mu_i + \mu'_i \geq 1  $ and zero otherwise.

\subsubsection{The thermodynamic limit and the extremal condition}
\label{2T-thermo}

Let us focus on the low temperature phase and take $ \beta > \beta_c $ and $ \beta' > \beta_c $ when  replica symmetry is broken. We proceed as we did in the single temperature case, by making  a similar set of three  assumptions on how to take the thermodynamic limit as in the single temperature case. As before  we  look for the minimum of $ A(r, \{\mu_i,\mu'_i\}) $ in (\ref{eq:A-two-T-dfn}) and  the only difference is that we now have the additional parameters $ n' $ and $ \{\mu'_i\} $. Using Lagrange multipliers we find that the minimum corresponds to  $ \beta \mu_i + \beta' \mu'_i $ being independent of $ i $. Summing on $ i $ and using the constraints $ \sum_{i=1}^{r} \mu_i = n $ and $ \sum_{i=1}^{r} \mu'_i = n' $ we find that 
\begin{equation} \label{eq:mu-mu-prime-constraint}
\beta \mu_i + \beta' \mu'_i = \frac{\beta n + \beta' n'}{r} \text{ for all } i.
\end{equation}
The value of $ r $ that gives the minimum  of 
$ A(r, \{\mu_i,\mu'_i\})= r f ({n \beta + n' \beta' \over r}) $ is then given by
\begin{equation}
f\left(\frac{\beta n + \beta' n'}{r}\right) 
- \frac{\beta n + \beta' n'}{r} f'\left(\frac{\beta n + \beta' n'}{r}\right) = 0.
\end{equation}
so that  from \eqref{eq:beta_c-eqtn} 
\begin{equation} \label{eq:r-2temp-dfnn}
r= \frac{\beta n + \beta' n'}{\beta_c}.
\end{equation}
Together with equation \eqref{eq:mu-mu-prime-constraint} this gives 
\begin{equation} \label{eq:mu-mu-prime-beta_c}
\beta \mu_i + \beta' \mu'_i = \beta_c,
\end{equation}
which constrains the fluctuations of $ \mu_i $ and $ \mu'_i $, but, unlike the single temperature case, (\ref{eq:mu-mu-prime-beta_c}) does not eliminate them completely.

One can however, without any further assumption,   recover (\ref{F10b}) from (\ref{eq:Ykk-mu-mu-average}). Using the replica form~\eqref{eq:Ykk-mu-mu-average} one can see that 
\begin{multline}
(k-n) \beta \left< Y_{k+1,k'} \right>_{n,n'} 
	+ (k'-n')\beta' \left< Y_{k,k'+1} \right>_{n,n'} = \\
	 \left< \vphantom{\frac{num}{den}} (\beta k + \beta' k' - \beta \mu_1 - \beta' \mu'_1) \right.\\
	 \times
	\left. r \; \frac{\mu_1 (\mu_1-1) \cdots (\mu_1 - k +1)}{n(n-1) \cdots (n-k+1)} 
	\frac{\mu'_1 (\mu'_1-1) \cdots (\mu'_1 - k' +1)}{n'(n'-1) \cdots (n'-k'+1)}
	\right>_{\mu_i,\mu'_i}.
\label{identity5}
\end{multline}
If we take the large $ N $ limit of the right hand side we expect (see(\ref{eq:mu-mu-prime-beta_c})) that the extremal condition $ \beta \mu_1 + \beta' \mu'_1 = \beta_c $ should apply. Then (\ref{identity5}) simplifies to give
\begin{equation} \label{eq:Y-replica-recurrance}
(k-n) \beta \left< Y_{k+1,k'} \right>_{n,n'}
+  (k'-n')\beta' \left< Y_{k,k'+1} \right>_{n,n'}
= (k \beta  + k'  \beta' - \beta_c ) \,\left< Y_{k,k'} \right>_{n,n'}
\end{equation}
the same recursion relation~\eqref{F10b} as in the direct calculation.
This gives at least some confidence in the assumptions that we have made in developing the replica approach so far for the two temperature problem. 

Ideally we would like to go a step further and recover the exact solution \eqref{F9a} directly from \eqref{eq:Ykk-mu-mu-average}  using the replica approach. The challenge is to find a way to compute the average over $ \{\mu_i,\mu_i'\} $ in \eqref{eq:Ykk-mu-mu-average} subject to the constraints \eqref{eq:r-2temp-dfnn} and \eqref{eq:mu-mu-prime-beta_c} that is valid when $ n,n' $ are no-longer integers. We have not found an approach that is sufficiently convincing to merit inclusion here. The problem essentially has to do with the $ r $ and the $ \{\mu_i,\mu_i'\} $ becoming non-integer. We will see however in section \ref{sec:Fluctuations} that by matching with the exact expressions of section  \ref{sec:direct-calculation} one can obtain the generating function of the $ \{\mu_i,\mu_i'\} $.

\section{Parisi overlap matrices}
\label{sec:Parisi-matrices}

For  the REM the replica method in section 3 could be implemented without explicitly using the replica overlap matrices. However, for more complex problems such as the Sherrington-Kirkpatrick model \cite{Sherrington_1975_Solvable} the saddle point equations are expressed in terms of replica overlap matrices and so it is useful to see what they look like in the case of the REM. In this section we describe the structure of these matrices in the single and two temperature case of the REM. The Parisi ansatz is used in the single temperature case and we show how it can be generalised to two temperatures in the case of the REM. 

We could have approached this by applying the replica method to the large $ p $ limit of the p-spin models introduced in \cite{Derrida_1981_Random} as was done in \cite{Gross_1984_simplest} for the single temperature case. However, the corresponding  two temperature calculation is rather long and is not essential to understanding how to generalise the Parisi ansatz.

\subsection{Single temperature case}

In the single temperature case $ \left\langle Z(\beta)^n \right\rangle $ is expressed in equation~\eqref{eq:Zn-sum-on-mu-exact-bis} as a sum over the parameters $ r $ and $ \mu_1,\mu_2, \ldots \mu_r $ with the constraint $ \sum_{i=1}^{r} \mu_i = n $. We can use these parameters to define an $n \times n $ replica overlap matrix  $ {\bf Q} (\{ \mu_i\}) $. We divide the $ n $ replicas into $ r $ groups of sizes $ \mu_1,\mu_2, \ldots \mu_r $. The overlap matrix is then defined as
\begin{equation} \label{eq:Qab-dfnn}
Q_{a,b} (\{ \mu_i \}) =
\begin{cases}
1, &\text{if replicas $ a,b $ are in the same group ,} \\
0, &\text{otherwise.}
\end{cases}
\end{equation}
This means that,  up to a permutation of the replica indices,  the $n\times n$ matrix  ${\bf  Q}(\{ \mu_i \}) $ consists of $ r $ blocks of size $ \mu_1 \times \mu_1, \mu_2 \times \mu_2, \ldots, \mu_r \times \mu_r, $  along   the diagonal where the matrix elements take value unity and they are zero elsewhere. For  example if $ n = 6, r =3, \mu_1=2, \mu_2=3, \mu_3 = 1 $ 
\begin{equation} \label{eq:Q-example}
{\bf Q}(\{ \mu_i \}) = \left(
\begin{array}{cc|ccc|cc|c}
{\bf 1} & {\bf 1} & 0 & 0 & 0 & 0  \\
{\bf 1} & {\bf 1} & 0 & 0 & 0 & 0 \\ \hline
0 & 0 & {\bf 1} & {\bf 1} & {\bf 1} & 0 \\
0 & 0 & {\bf 1} & {\bf 1} & {\bf 1} & 0 \\
0 & 0 & {\bf 1} & {\bf 1} & {\bf 1} & 0 \\ \hline
0 & 0 & 0 & 0 & 0 & {\bf 1} \\
\end{array}
\right) \ . 
\end{equation}
(Here we have taken $ Q_{a,a} = 1 $ for simplicity). In terms of this overlap matrix, one can rewrite 
(\ref{eq:A-single-T-dfnn-nu})   as
\begin{equation}\label{eq:A-single-T-dfnn-Qab}
A(r, \{\mu_i\}) = r \log 2 + \frac{(\beta J)^2}{4} 
\sum_{a=1}^{n} \sum_{b=1}^{n} Q_{a,b}(\{ \mu_i \}).
\end{equation}
where we have used that  $ \sum_{a=1}^{n} \sum_{b=1}^{n} Q_{a,b}(\{ \mu_i \}) = \sum_{i=1}^{r} \mu_i^2 $. 

The thermodynamic limit gives us the extremal condition~\eqref{eq:mu-a-equal-mu} which indicates that the dominant form of ${\bf  Q}(\{ \mu_i \}) $ has all the $ \mu_i $ equal. This fixed block structure gives the one step RSB form of the overlap matrices introduced by Parisi \cite{Parisi_1979_Infinite} to solve mean field spin glass models such as the Sherrington-Kirkpatrick model~\cite{Sherrington_1975_Solvable}. It should be noted that any permuation of the replica indices will also give a ${\bf  Q}(\{ \mu_i \}) $ that satisfies the extremal condition and we should sum over all these saddle points when computing physical properties such as $ P(q) $  (see \cite{DeDominicis_1983_Weighted}).

\subsection{Two temperature case}

In the two temperature case $ \left\langle Z(\beta)^n Z(\beta')^{n'} \right\rangle $ is expressed in equation~\eqref{eq:ZZnn-sum-on-mu-exact} as a sum over the parameters $ r $; $ \mu_1,\mu_2, \ldots \mu_r $ and $ \mu'_1,\mu'_2, \ldots \mu'_r $ with the constraints $ \sum_{i=1}^{r} \mu_i = n $ and  $ \sum_{i=1}^{r} \mu'_i = n' $. We can use these parameters to define three different replica overlap matrices.  We divide the $ n $ replicas into $ r $ groups of size $ \mu_1,\mu_2, \ldots \mu_r $ and the $ n' $ replicas into $ r $ groups of size $ \mu'_1,\mu'_2, \ldots \mu'_r $. We then have the single temperature $n \times n $ replica overlap matrix  $ {\bf Q} (\{ \mu_i\}) $ defined in equation~\eqref{eq:Qab-dfnn} and the equivalent $n' \times n' $ matrix $ {\bf Q'} (\{ \mu'_i\}) $ at inverse temperature $ \beta' $. We can also define an $n\times n'$ overlap matrix $ {\bf R} (\{ \mu_i,\mu'_i\}) $ between the inverse temperature $ \beta $ and the inverse temperature $ \beta' $ by
\begin{equation} \label{eq:Rab-dfnn}
R_{a,b'}(\{\mu_i,\mu'_i\})=
\begin{cases}
1, &\text{if replicas $ a,b' $ are in the same group;} \\
0, &\text{otherwise.}
\end{cases}
\end{equation}
So all the matrix elements of the rectangular matrix $ {\bf R} (\{ \mu_i , \mu'_i \}) $ are zero except  $ r $ blocks of sizes $ \mu_1 \times \mu'_1, \mu_2 \times \mu'_2, \ldots, \mu_r \times \mu'_r, $  along  the diagonal where they take value unity. It has the property that $ \sum_{a=1}^{n} \sum_{b'=1}^{n'} R_{a,b'}(\{ \mu_i , \mu'_i \}) = \sum_{i=1}^{r} \mu_i \mu'_i $ so that we can write \eqref{eq:A-two-T-dfn} as
\begin{multline} \label{eq:A-two-T-dfnn-QPQ}
A\big(r, \{\mu_i, \mu'_i\}\big) = r \log 2 
+ \frac{(\beta J)^2}{4} 
\sum_{a=1}^{n} \sum_{b=1}^{n} Q_{a,b}(\{ \mu_i \})\\
+ \beta \beta' \frac{J^2}{2} 
\sum_{a=1}^{n} \sum_{b'=1}^{n'} R_{a,b'}(\{ \mu_i, \mu'_i \})
+ \frac{(\beta' J)^2}{4} 
\sum_{a'=1}^{n'} \sum_{b'=1}^{n'} Q'_{a',b'}(\{ \mu'_i \}).
\end{multline}

As an example of the overall matrix, 
for  $ n = 6,n'=9, r =3, \mu_1=2, \mu_2=3, \mu_3 =  1, \mu'_1=4, \mu'_2 = 1, \mu'_3 = 3$,
one has
\begin{equation} \label{eq:QR-example}
\left(
\begin{array}{lll}
\mathbf{Q} & \mathbf{R} \\
\mathbf{R}^T & \mathbf{Q}'
\end{array}
\right)
= \left(
\begin{array}{cc|ccc|c||cccc|c|ccc}
{\bf 1} & {\bf 1} & 0 & 0 & 0 &  0  &  {\bf 1} & {\bf 1} & {\bf 1} & {\bf 1}  & 0 & 0 & 0 & 0  \\
{\bf 1} & {\bf 1} & 0 & 0 & 0 &  0  &  {\bf 1} & {\bf 1} & {\bf 1} & {\bf 1}  & 0 & 0 & 0 & 0  \\
\hline
0 & 0 & {\bf 1} & {\bf 1} & {\bf 1} &  0  & 0 & 0 & 0 & 0 & {\bf 1}  & 0 & 0 & 0  \\
0 & 0 & {\bf 1} & {\bf 1} & {\bf 1} &  0  & 0 & 0 & 0 & 0 & {\bf 1}  & 0 & 0 & 0  \\
0 & 0 & {\bf 1} & {\bf 1} & {\bf 1} &  0  & 0 & 0 & 0 & 0 & {\bf 1}  & 0 & 0 & 0  \\
\hline
0 & 0 & 0 & 0 & 0 &  {\bf 1}  & 0 & 0 & 0 & 0 & 0 &  {\bf 1} & {\bf 1} & {\bf 1}  \\
\hline
\hline
{\bf 1} & {\bf 1} & 0 & 0 & 0 &  0  & {\bf 1} & {\bf 1} & {\bf 1} & {\bf 1}  & 0 & 0 & 0 & 0  \\
{\bf 1} & {\bf 1} & 0 & 0 & 0 &  0  & {\bf 1} & {\bf 1} & {\bf 1} & {\bf 1}  & 0 & 0 & 0 & 0  \\
{\bf 1} & {\bf 1} & 0 & 0 & 0 &  0  & {\bf 1} & {\bf 1} & {\bf 1} & {\bf 1}  & 0 & 0 & 0 & 0  \\
{\bf 1} & {\bf 1} & 0 & 0 & 0 &  0  & {\bf 1} & {\bf 1} & {\bf 1} & {\bf 1}  & 0 & 0 & 0 & 0  \\
\hline
0 & 0 & {\bf 1} & {\bf 1} & {\bf 1} &  0  & 0 & 0 & 0 & 0 & {\bf 1} &  0 & 0 & 0  \\
\hline
0 & 0 & 0 & 0 & 0 &  {\bf 1}  & 0 & 0 & 0 & 0 & 0  & {\bf 1} & {\bf 1} & {\bf 1}  \\
0 & 0 & 0 & 0 & 0 &  {\bf 1}  & 0 & 0 & 0 & 0 & 0  & {\bf 1} & {\bf 1} & {\bf 1}  \\
0 & 0 & 0 & 0 & 0 &  {\bf 1}  & 0 & 0 & 0 & 0 & 0  & {\bf 1} & {\bf 1} & {\bf 1}  \\
\end{array}
\right)
\end{equation}

This type of two temperature order parameter has already been discussed in the context of spin models in a number of works on temperature chaos (see \cite{Rizzo_2009_Chaos} for a review).
 
\subsection{When the numbers \texorpdfstring{$n$ and $n'$}{n and n'} of replicas become non integer}

In the thermodynamic limit, in the case of a single temperature, we have seen in section \ref{1T-thermo}  
that the number $r$ of blocks is fixed and that all the $\mu_i$ are equal  to the value $\mu={\beta_c \over \beta}$  (see (\ref{eq:r-and-mu-1T})).  Therefore the matrix ${\mathcal{Q}}$ in 
(\ref{eq:Q-example}) takes precisely the form first proposed by Parisi \cite {Parisi_1979_Infinite} with blocks  of equal sizes along the diagonal.

In the case of two temperatures, we have seen in section \ref{2T-thermo} that the number $r$ of blocks is still fixed  (see (\ref{eq:r-2temp-dfnn})) and that there is a constraint $\beta \mu_i + \beta' \mu_i' = \beta_c$ (see (\ref{eq:mu-mu-prime-beta_c})) for each pair $\mu_i,\mu_i'$. The simplest assumption would be to take $\mu_i=\mu$ and $\mu_i'=\mu'$ independent of $i$. This choice is not consistent with the exact expressions (\ref{F7a})  of $\langle Y_{k,k'} \rangle $ presented in section \ref{sec:direct-calculation} and therefore $\mu_i$ and $\mu_i'$ fluctuate subject to the constraint (\ref{eq:mu-mu-prime-beta_c}).

However, to obtain the exact results \eqref{F7a} or \eqref{F9a}, using the replica method we must sum over all the saddle points that satisfy the constraints \eqref{eq:r-2temp-dfnn} and \eqref{eq:mu-mu-prime-beta_c}. In the single temperature case this was fairly straightforward (see \cite{DeDominicis_1983_Weighted}) because all the saddle points are related by a simple permutation of the replica indices. In the two temperature case this is no longer true, as discussed in detail in \cite{Rizzo_2002_Ultrametricity}, and it is not clear how to sum over the saddle points.

\section[The fluctuations in block sizes]{The fluctuations in block sizes \texorpdfstring{$ \mu_i, \mu'_i $}{}}
\label{sec:Fluctuations}

In this section we  analyse the fluctuations of the  block sizes $ \mu_i, \mu'_i $ in the thermodynamic limit. We first obtain the mean and the variance of the $\mu_i$ and $\mu_i'$.  We will then
obtain the  moment generating function for the distribution of these block sizes  from the exact expression~\eqref{F9a} for $ \langle Y_{k, k'}\rangle_{n,n'} $. One outcome of our  results is that the $\mu_i$ and the $\mu_i'$ do fluctuate even in the equal temperature  case. However  extracting the distribution of $P(\mu_i,\mu_i')$ from this generating function is not an easy task and  can be interpreted as a   signed measure,  i.e.  a measure with negative probabilities. Also not all properties of the distribution of these block sizes have physical implications: for example,  we will see  that in the limit $n\to 0^- $ and $n' \to 0^-$ these distributions depend on the ratio $n'/n$ although all physical properties have a limit independent of this ratio.  

In this section $ n,n' $ are negative real numbers because our analysis is based on the exact expression \eqref{F9a} which is only valid for this range of values.

\subsection{The first moments \texorpdfstring{of $\mu_i$}{}}

As $Y_{1,0}=1 $ (see \eqref{eq:y-kk-dfnn}) and as  $r$ 
does not fluctuate (see  (\ref{eq:r-2temp-dfnn}))
one can show from
 (\ref{eq:Ykk-mu-mu-average}) that     
\begin{align}
\langle \mu_i \rangle_{\{\mu_i,\mu'_i\}} &= {n \beta_c \over n \beta + n' \beta'} \notag \\
\langle \mu_i^2 \rangle_{\{\mu_i,\mu'_i\}} &= 
{n \beta_c(1-(1-n) \langle Y_{2,0} \rangle_{n,n'})    \over n \beta + n' \beta'}
= {n \beta_c ( \beta_c -n' \beta' \langle  Y_{1,1}\rangle_{n,n'} )    \over \beta(n \beta + n' \beta')}
\label{low-moments}
\end{align}
where we have used the  relation between $\langle Y_{2,0} \rangle_{n,n'} $ and $\langle Y_{1,1} \rangle_{n,n'} $ 
$$\langle Y_{2,0}\rangle_{n,n'} = {\beta-\beta_c + n' \beta' \langle Y_{1,1}\rangle_{n,n'}  \over (1-n) \beta}$$
which follows from  (\ref{F10b}) and the fact that $Y_{1,0}=1$.

From these expressions (\ref{low-moments}) one can notice first that the limit of the first moment  $\langle \mu_i \rangle $,  when $n \to 0^-$ and $n' \to 0^-$,  depends on  the ratio $n'/n$. This means that not all the properties of the $\mu_i$ have a physical meaning, since one expects all physical properties to be independent of this ratio when $n$ and $n'$ vanish.

One can also notice that the variance of $\mu_i$ is in general non-zero. Depending on $n,n',\beta,\beta'$, this variance may change its sign, implying that the distribution of $\mu_i$ is not really a probability distribution. For example when $\beta=\beta'$, one has \eqref{F9b} 
$$\langle Y_{1,1}\rangle_{n,n'}= \langle Y_{2,0} \rangle_{n,n'} 
= \frac{1 - {\beta_c \over \beta}}{1-n-n'}$$ 
and 
$$\langle \mu_i^2 \rangle_{\{\mu_i,\mu'_i\}} - \langle \mu_i \rangle_{\{\mu_i,\mu'_i\}}^2 
= {n n' \beta_c (  \beta(n+n') - \beta_c) \over \beta^2 (n+n')^2 (n+n'-1)}
$$
which in general does not vanish and can be of either sign.
We already observed such negative variances of the block sizes   when we tried to reproduce, using replicas, finite size corrections of the REM at a single temperature \cite{Derrida_2015_Finite}.

Expressions  similar to (\ref{low-moments}) can be obtained for $\mu_i'$ by using either the symmetry $n,\beta \leftrightarrow n' \beta'$ or the fact that
 the sum $\beta \mu_i + \beta' \mu_i'=\beta_c$  does not fluctuate 
   (\ref{eq:mu-mu-prime-beta_c}).   

\subsection{The generating function \texorpdfstring{of $\mu_i$ and $\mu_i'$}{}}
We are now going to obtain the exact expression of the generating function 
$\left\langle x^{\mu_1} y^{\mu'_1} \right\rangle_{\{\mu_i,\mu'_i\}} $
where the  average  $ \left\langle \cdot \right\rangle_{\{\mu_i,\mu'_i\}} $ is defined in equation~\eqref{eq:F-mu-mu-average}. Taking the Taylor expansions of $ x^{\mu_1} , y^{\mu'_1} $ about $ x=1, y=1 $ the generating function  can be written as 
\begin{multline} \label{eq:xx-expansion}
\left\langle x^{\mu_1}  y^{\mu'_1} \right\rangle_{\{\mu_i,\mu'_i\}} 
	= \sum_{k \ge 0} \sum_{k' \ge 0}
	\frac{(x-1)^k}{k!} \frac{( y-1)^k}{k'!} \\
	\times
	\left\langle \mu_1 (\mu_1-1) \cdots (\mu_1 - k +1)
		\mu'_1 (\mu'_1-1) \cdots (\mu'_1 - k' +1)
		\right\rangle_{\{\mu_i,\mu'_i\}}.
\end{multline}
In the thermodynamic limit we can express the average on $ \{\mu_i,\mu'_i\} $ on the right hand side in terms of $ \left< Y_{k,k'} \right>_{n,n'} $ using equation~\eqref{eq:Ykk-mu-mu-average}. This gives
\begin{multline} \label{eq:Ykk-mu-mu-average-limit}
\left<\mu_1 (\mu_1-1) \cdots (\mu_1 - k +1) \quad
\mu'_1 (\mu'_1-1) \cdots (\mu'_1 - k' +1)
\right>_{\{\mu_i,\mu'_i\}}\\
	= \frac{1}{r} n'(n'-1) \cdots (n'-k'+1)
		n(n-1) \cdots (n-k+1) \left< Y_{k,k'} \right>_{n,n'}
\end{multline}
where we have  used the fact that $ r $ does not fluctuate in the thermodynamic limit (see (\ref{r-expressiona})). Using the exact expression~\eqref{F9a} for $ \left< Y_{k,k'} \right>_{n,n'} $ we obtain
\begin{multline}
\left<\mu_1 (\mu_1-1) \cdots (\mu_1 - k +1) \quad
	\mu'_1 (\mu'_1-1) \cdots (\mu'_1 - k' +1)
	\right>_{\{\mu_i,\mu'_i\}}\\
	=
	(-1)^{k+k'}
	\frac{ \int_0^\infty dv  \ v^{k'-n' -1}
	\Psi\left( k + k' {\beta' \over \beta} ; v  \right)
	\ (-\psi(v))^{r-1}
	}{
	\int_0^\infty dv   \ v^{-n'-1}\
	(-\psi(v))^r }
\end{multline}
Finally, substituting into equation~\eqref{eq:xx-expansion} and summing on $ k,k' $ we obtain
\begin{equation} \label{eq:xx-exact}
\left\langle x^{\mu_1}  y^{\mu'_1} \right\rangle_{\{\mu_i,\mu'_i\}} 
	= 
	\frac{x^{\frac{\beta_c}{\beta}}  \int_0^\infty dv  \ v^{-n' -1}
		\left(-\psi\left( v  y x^{-\frac{\beta'}{\beta} }  \right)\right)
		\ (-\psi(v))^{r-1}
	}{
		\int_0^\infty dv   \ v^{-n'-1}\
		(-\psi(v))^r }.
\end{equation}
where $ r $ is given by \eqref{r-expressiona}.  (Note that, as mentioned at the beginning of this section the above expression \eqref{eq:xx-expansion} is valid for $n<0$ and $n'<0$.)  

\begin{remark}
	{\it One recovers \eqref{eq:mu-mu-prime-beta_c}}\\*
	Making the substitution $ x= z^\beta,  y=z^{\beta'} $ in \eqref{eq:xx-exact} gives
	\begin{equation} 
	\left\langle z^{\beta \mu_1 + \beta' \mu'_1}
	\right\rangle_{\{\mu_i,\mu'_i\}} 
	= z^{\beta_c}.
	\end{equation}
	This confirms the fact that the sum  $ \beta \mu_1 + \beta' \mu'_1 $ does not fluctuate and takes the value $ \beta_c $,  as expected from   \eqref{eq:mu-mu-prime-beta_c}.
\end{remark}

\begin{remark}
	{\it The $n,n' \to 0^- $ limit}\\*
	One can show using the asymptotics (\ref{asymptotics}) that, in the limit $n\to 0^-$ and $n' \to 0^-$,
	the generating function (\ref{eq:xx-exact}) becomes
	
	\begin{align}
		\left\langle x^{\mu_1}  y^{\mu'_1} \right\rangle_{\{\mu_i,\mu'_i\}}
		= &
		{\beta n x^{\beta_c\over \beta} + \beta' n' y^{\beta_c \over \beta'} \over \beta n + \beta' n'} + 
		{\beta \beta' n  n' ( y^{\beta_c \over \beta'}-
			x^{\beta_c\over \beta}
			)  \over \beta_c(\beta n + \beta' n') } \log \left( {  \Gamma\left(1- {\beta_c \over \beta'}\right) \over  \Gamma\left(1- {\beta_c \over \beta}\right) }  \right)
		\nonumber
		\\ & + 
		{\beta  n  n'   \over \beta n + \beta' n' } x^{\beta_c \over \beta} \int_0^\infty \log v {d \over dv} \left[ {\psi(y x^{-{\beta'\over \beta}}  v)\over \psi( v)}   \right]  dv
		\label{n=0-n'=0}
		\\ & +  o(n,n')
		\nonumber
	\end{align}
	To leading order one finds that the distribution of $\mu_1$ and $\mu_1'$ consists of two delta functions
	
	$$P(\mu_1,\mu_1') = 
	\left[
	{n \beta   \over n \beta  + n' \beta' }
	\delta\left( \mu_1- {\beta_c\over \beta}\right) 
	+ {n' \beta'   \over n \beta  + n' \beta' }
	\delta\left( \mu_1 \right) \right] \ \beta' \delta( \beta \mu_1 + \beta' \mu_1' - \beta_c) 
	$$
	Clearly this expression does not contain any information on the overlaps $\langle Y_{k,k'}\rangle$. In fact the generating function of these overlaps  only appears in the first order term in (\ref{n=0-n'=0}). Note also that as for the variance of $\mu_1$, the $n\to 0^-, n'\to 0^-$ limit depends on the ratio $n'/n$. 
\end{remark}

\subsection{Trying to describe \texorpdfstring{$P(\mu_1,\mu_1')$}{the probability distribution}}

As the variance of the $\mu_i$ can become negative, it is clear from the very beginning that it is not possible to find a meaningful distribution of the block sizes compatible with the generating function
(\ref{eq:xx-exact}).  We made a number of attempts which became rather complicated and we don't think it is of much interest to mention them here. Let us however discuss briefly  one case for which we could get a rather simple picture, the equal temperature case, i.e.  when $\beta=\beta'$.

In this case we  have an explicit expression (\ref{equalT1}) of the function $\psi(v)$. 
Then (\ref{eq:xx-exact}) becomes
$$
\left\langle x^{\mu_1}  y^{\mu'_1} \right\rangle_{\{\mu_i,\mu'_i\}}
= {\int_0^\infty v^{-n'-1} \ 
( 1+ v )^{n+n'-{\beta_c\over \beta} }
( x+ v y)^{\beta_c\over \beta}  dv
\over 
 \int_0^\infty v^{-n'-1} \ 
( 1+ v )^{n+n'} 
dv
} $$
After a simple change of variable
$v=(1-t)/t$
this  becomes

\begin{equation}
\left\langle x^{\mu_1}  y^{\mu'_1} \right\rangle_{\{\mu_i,\mu'_i\}}
= {\Gamma(-n-n') \over \Gamma(-n) \, \Gamma(-n')} \int_0^1  dt \  t^{-1-n} (1-t)^{-1-n'} \Big(t x + (1-t)
 y\Big)^{\beta_c \over \beta}
\label{pseudo-binomial}
\end{equation}

In order to give an interpretation to (\ref{pseudo-binomial}) let us consider a random variable  $s$,  sum of $m$ i.i.d. random variables  $\tau_i$ which take the value $\tau_i=1$ with probability $t$ and $\tau_i=0$ with probability $1-t$. 
The distribution of $s$ is a binomial distribution and one has
\begin{equation}
 \langle z^s\rangle= (z t + 1-t)^m \ . 
\label{binomial}
\end{equation}
Let us further  consider that the parameter $t$ is itself randomly distributed according to some distribution $\rho(t)$ so that the distribution of $s$ becomes a superposition of binomial distributions.
Then the generating function of $s$ becomes
$$\langle z^s\rangle = \int_0^1 \rho(t) (z t +1-t)^m dt \ . $$
This is exactly the form we have in  (\ref{pseudo-binomial}) (by taking $x=z $ and $y=1$) if one chooses for $\rho(t)$ 
$$
\rho(t)= {\Gamma(-n-n') \over \Gamma(-n) \, \Gamma(-n')}   t^{-1-n} (1-t)^{-1-n'}
$$
(remember that here $n$ and $n'$ are negative).

Therefore the distribution of $\mu_1$ can be thought as a superposition of binomial distributions. The only odd aspect is that $s$ is a sum of $m={\beta_c \over \beta}$ binary variables,  that is $s$ is a sum of a non-integer number of random  variables! 

\begin{remark}
	{\it A signed measure}\\*
	If one takes a non-integer $m$ in (\ref{binomial}) one gets by expanding in powers of $z$
	$$\langle z^s\rangle = \sum_{p=0} (1-t)^{m-p} t^p \  {m(m-1) \cdots (m-p+1) \over p!} \ z^p$$
	which one can  interpret,  for $ t < \frac{1}{2} $, as the probability $P(s)$ of $s$   being a signed measure concentrated on positive integers. Expanding in powers of $1/z$ leads, for ${1 \over 2} < t < 1$,  to a different signed measure. Combining these two representations by cutting the integral (\ref{pseudo-binomial}) into two parts ($t<{1\over 2}$ and $t>{1\over 2}$) leads to a signed measure concentrated on the points 
	$(\mu_i=p, \mu_i'={\beta_c\over \beta}-p)$ and
	$(\mu_i={\beta_c\over \beta}-p,
	\mu_i'=p)$ for all positive integers $p\ge0$. 
\end{remark}

\section{Conclusion}

In this paper we have analysed the distribution of overlaps \eqref{eq:Pq-beta-beta} between two copies of the same REM at two temperatures. A direct calculation was used to obtain exact expressions \eqref{F9a} for the two temperature overlaps \eqref{eq:y-kk-dfnn} in the thermodynamic limit. Generalising this approach allows us to quantify \eqref{variance} the non-self-averaging effects illustrated in Figure~\ref{Fig1}.

An alternative approach using the replica method enables us to obtain expressions for the two temperature overlaps in terms of replicas \eqref{eq:Ykk-mu-mu-average}. In the thermodynamic limit the exact and replica expressions satisfy the same Ghirlanda-Guerra type recurrence relation, \eqref{F10b} and \eqref{eq:Y-replica-recurrance}, giving confidence that the replica expressions are valid. We also proposed a way to generalise the Parisi ansatz \eqref{eq:QR-example}, in the one step RSB form, to the two temperature case which is consistent with the replica expressions for the overlaps. In contrast to the single temperature case we find that the block sizes at the two different temperatures fluctuate even in the thermodynamic limit subject the constraint \eqref{eq:mu-mu-prime-beta_c}. We characterised these fluctuations in terms of a moment generating function for the block sizes \eqref{eq:xx-exact}. It is well known that in the single temperature case the strange properties of Parisi's RSB ansatz  (non-integer block sizes and number of blocks) lead to a perfectly good physical description in terms of overlaps \cite{Mezard_1984_Nature}.  In the two temperature case the distribution of $ \mu_i $ and $ \mu_i' $ analysed in section~\ref{sec:Fluctuations} also has strange properties, but it remains an open question as to which of these properties have a clear physical interpretations.

It would be interesting to extend both the exact and replica approaches to the generalised random energy model \cite{Derrida_1985_Generalization}, directed polymer in a random medium \cite{Derrida_1988_Polymers} and other models where exact methods are likely to be tractable. 
In contrast to the single temperature case, the multi-temperature overlaps should be different in the REM and in the directed polymer problem on a tree because the lowest energies of the directed polymer can be thought of as a decorated Poisson process and it has been proved that the multi-temperature overlaps depend on the decoration \cite{Pain_2018_Two}. 
One could also look at spin models where one step RSB occurs to see if the two temperature ansatz with fluctuating block sizes is applicable. An obvious starting point would be the p-spin spherical model proposed in \cite{Crisanti_1992_sphericalp}. In order to address these spin problems, where exact expressions for the two temperature overlaps are not currently available, it would be essential to develop a systematic approach to summing over the fluctuations in block sizes in the replica expressions. 

\appendix

\section{Direct calculation of the overlaps (\ref{F7a},\ref{F9a})}
\label{sec:appendix-direct calculation}

To begin with, it is easier  to think that the energies can take only a discrete set of values $E_\nu$ indexed by  $\nu$ and that the partition function at inverse temperature $\beta$ is given by
$$ Z(\beta) = \sum_{\nu} y_\nu \,  e^{-\beta E_\nu} $$
where 
$$y_\nu = \left\{  \begin{array}{ccc}
1 & \text{with probability} & p_\nu \\
0 & \text{with probability} &1 -p_\nu 
\end{array}
\right.  \ . 
$$ 
So a given sample is specified by the value of all these binary random variables $y_\nu$.
Then  the probability of finding $k$ copies at temperature $\beta$ and $k'$ copies at temperature $\beta'$ in  the same configuration is given by 
$$Y_{k, k'} = {\sum_{\nu } \, y_\nu \,     e^{
 -(\beta k + \beta' k')E_\nu
}
\over Z(\beta)^{k} \, Z(\beta')^{k'} } $$
These  
$Y_{k, k'}$ are random quantities as they depend on the realization of the $y_\nu$'s.
Using the identity  $Z^{-k} = \Gamma(k)^{-1}  \int_0^\infty dt \, e^{-t Z} \, t^{k-1}$ one gets
$$\begin{aligned} Y_{k,k'} = & 
\sum_{\nu }   \, y_\nu \,     e^{
 -(\beta k + \beta' k')E_\nu}
\\ 
& \int_0^\infty {t^{k-1} dt  \over \Gamma(k)}
\int_0^\infty {t'^{k'-1} dt'  \over \Gamma(k')}
\exp\left[- \sum_{\nu'} y_{\nu'} \Big( t e^{-\beta E_{\nu'} } + t'  e^{-\beta' E_{\nu'} } \Big) \right] \end{aligned} $$
Averaging over the  $y_\nu$'s leads to
$$\begin{aligned} \langle Y_{k,k'} \rangle 
=  
 \int_0^\infty {t^{k-1} dt \over  \Gamma(k)}
\int_0^\infty {t'^{k'-1} dt' \over  \Gamma(k')}
 & \sum_{\nu }   \, p_\nu \,       e^{
 -(\beta k + \beta' k')E_\nu}
\ \exp\left[ -  
 t  e^{-\beta E_{\nu} }    
- t'  e^{-\beta' E_{\nu} }  \right] 
\\  & 
  \times  \prod_{\nu' \neq \nu}    \Big( 1- p_{\nu'} + p_{\nu'}
\exp\left[ - t e^{-\beta E_{\nu'} } - t'  e^{-\beta' E_{\nu'} } \right] \Big) \end{aligned} $$

Now if we go to the continuum limit, by saying that each energy interval $(E,E+dE)$  is either occupied by an energy level or empty and if we choose as in (\ref{poisson})
$$p_\nu =  C'  e^{\beta_c E}  \, dE \ \ \ \ \  {\rm with} \ \ \ \ \ C'= C e^{-\beta_c E_0} $$
 one gets 
$$ \langle Y_{k,k'} \rangle 
=  
 \int_0^\infty {t^{k-1} dt \over  \Gamma(k)}
\int_0^\infty {t'^{k'-1} dt' \over  \Gamma(k')}
 \ W(k,k';t,t') 
 \  e ^{ w(t,t')} 
$$
where
$$w(t,t')= C'   \int e^{\beta_c E} dE  \Big( 
\exp\left[ - t e^{-\beta E}  - t'  e^{-\beta' E } \right]  - 1 \Big)
$$
and 

$$W(k,k'; t,t') =  C' \int e^{\beta_c E} dE  \ 
e^{ 
 -(\beta k + \beta' k')E}
\exp\left[ - t e^{-\beta E}  - t'  e^{-\beta' E } \right] 
$$
Then these expressions can be simplified by noticing that
$$w( t,t') =  {C' \over \beta}\  t^{{\beta_c \over \beta} } \ 
 \psi\left({t' \over t^{\beta'\over \beta}}\right) $$ and 
$$W( k,k'; t,t') =  {C' \over \beta}\  t^{{\beta_c \over \beta} -k - {\beta' \over \beta} k'} \ 
\Psi\left( k + k' {\beta' \over \beta} 
\, ; \, 
{t' \over t^{\beta'\over \beta}}
 \right) $$ where
\begin{equation}
\psi(v) = \int_0^\infty  du 
 \ ( e^{-u - v u^{\beta' \over \beta}}-1)
\ 
 u^{-1-{\beta_c \over \beta}}
\label{psi-defi}
\end{equation}
and
\begin{equation}
\Psi( z \, ; \, v ) = \int_0^\infty  du \
 e^{-u - v u^{\beta' \over \beta}} \ 
  u^{z-1 -{\beta_c \over \beta}} 
\label{Psi-defi}
\end{equation}
This leads to
\begin{equation}
\begin{aligned}
 \langle Y_{k,k'} \rangle 
= 
{\beta \over \beta_c}
\ 
  {1   \over \Gamma(k)}\  {1 \over \Gamma(k')}
  \int_0^\infty dv  \ v^{k'-1}
{\Psi\left( k + k' {\beta' \over \beta} ; v  \right)
\over \big(-\psi(v)\big)}
\end{aligned}
\label{F7}
\end{equation}

\begin{remark}
	To  generalize (\ref{F7}) one can define  the probability of finding $k_1$ copies at inverse temperature $\beta$ and $k_1'$ copies of the same system  at inverse temperature $\beta'$ in  the same configuration, and similarly  $k_2$ and $k_2'$ in a different configuration and so on  
	i.e.
	$$
	Y_{k_1,k_1'; \cdots k_p,k_p'} = {  {\sum_{\nu_1 \cdots \nu_p} e^{
				-(\beta  k_1  + \beta' k_1') E_{\nu_1}- \cdots 
				(\beta  k_p  + \beta' k_p') E_{\nu_p} } \over Z(\beta)^{k_1+ \cdots k_p} \ Z(\beta')^{k_1'+ \cdots k_p'} }} $$
	where, in the sum,  the configurations $\nu_1 \neq \nu_2 \neq \cdots \nu_p$  are all different. By a straightforward extension of the above calculation one gets
	\begin{equation}
		\begin{aligned}
			&  \langle Y_{k_1,k_1'; \cdots k_p,k_p'} \rangle 
			= 
			{\beta \over \beta_c}   \Gamma (p) 
			\ 
			%  {1   \over \Gamma(k_1+k_2+ \cdots k_p )}\  {1 \over \Gamma(k_1'+k_2' + \cdots k_p')
			{1   \over \Gamma(k_1+ \cdots k_p )}\  {1 \over \Gamma(k_1' + \cdots k_p')
			}
			% \\  &  \times  \int_0^\infty dv  \ v^{k_1'+k_2' + \cdots k_p'-1} \ 
			\\  &  \times  \int_0^\infty dv  \ v^{k_1' + \cdots k_p'-1} \ 
			{\Psi\left( k_1 + k_1' {\beta' \over \beta} ; v  \right)
				% \
				% \Psi\left( k_2 + k_2' {\beta' \over \beta} ; v\right)
				\cdots
				\Psi\left( k_p + k_p' {\beta' \over \beta} ; v\right)
				\over \big(-\psi(v)\big)^p}
		\end{aligned}
		\label{F8}
	\end{equation}
\end{remark}

\begin{remark}
	In the way the above formulae are written, $\beta$ and $\beta'$ seem to play asymmetric roles. One can however check from the definitions (\ref{psi-defi},\ref{Psi-defi}) of $\psi$ and $\Psi$ that
	\begin{equation}
		\psi_{\beta,\beta'}(v) = {\beta \over \beta'} \ v^{\beta_c \over \beta'}  \ \psi_{\beta',\beta} ( v^{-{\beta \over \beta'}} )
		\label{psi1}
	\end{equation} 
	\begin{equation}
		\Psi_{\beta,\beta'}(z \ ; \ v) = {\beta \over \beta'} \ v^{{\beta_c \over \beta'}-{\beta \over \beta'} z }  \ \Psi_{\beta',\beta} \left({\beta \over \beta'} z \  ; \  v^{-{\beta \over \beta'}} \right) 
		\label{Psi1}
	\end{equation} 
	and using these relations one can easily prove that the expressions (\ref{F7})  and (\ref{F8}) 
	are left unchanged by  the symmetry
	$$\Big(\beta, \beta', \{k_1,\cdots k_p \}, \{k_1', \cdots k_p'\} \Big)
	\longleftrightarrow 
	\Big(\beta', \beta, \{k_1',\cdots k_p' \}, \{k_1, \cdots k_p\}\Big) $$
\end{remark}

\begin{remark}
	When $\beta=\beta'$, the expressions   (\ref{psi-defi}) and (\ref{Psi-defi}) become
	\begin{equation}
		\psi(v) = \Gamma\left(-{\beta_c \over \beta} \right) \ (1+v)^{\beta_c \over \beta} 
		\label{equalT1}
	\end{equation}
	\begin{equation}
		\Psi(v) = \Gamma\left(z -{\beta_c \over \beta} \right) \ (1+v)^{{\beta_c \over \beta}-z} \ . 
		\label{equalT2}
	\end{equation}
	The integrals in 
	(\ref{F8})  can then be performed and one gets
	\begin{equation}
		\begin{aligned}
			\langle Y_{k_1,k_1';  \cdots k_p,k_p'} \rangle
			& =
			{\beta \over \beta_c}{   \Gamma (p) 
				\over \Gamma(k_1+k_1'+ 
				% k_2+  k_2'+ 
				\cdots k_p+k_p' )}\   \prod_{i=1}^p \left( {\Gamma (k_i+k_i'-{\beta_c\over \beta}) \over -\Gamma(-{\beta_c \over \beta}) } \right)
		\end{aligned}
		\label{F8a}
	\end{equation}
	which was already known (see for example \cite{Derrida_2015_Finite}).
\end{remark}

\begin{remark}
	It is easy to show, using an integration by parts in (\ref{Psi-defi}), that (for $z>{\beta_c \over \beta}$)
	$$\left(z-{\beta_c \over \beta} \right) \Psi(z \ ; \ v)
	= \Psi(z+1 \ ; \ v)
	+ {\beta' \over \beta} \ v \  \Psi \left(z+{\beta' \over \beta}\ ; \ v  \right) 
	$$
	and this leads (see (\ref{F8})) to relationships between the 
	$ \langle Y_{k_1,k_1'; \cdots k_p,k_p'} \rangle $
	\begin{equation}
		\begin{aligned} \left( k_1 \beta + k_1'\beta' - \beta_c  \right) 
			\langle Y_{k_1,k_1'; \cdots k_p,k_p'} \rangle  =  &
			(k_1 + \cdots k_p)\beta \,  \langle Y_{k_1+1,k_1'; \cdots k_p,k_p'} \rangle  
			\\ & +  (k_1' + \cdots k_p')\beta' \, \langle Y_{k_1,k_1'+1; \cdots k_p,k_p'} \rangle  
		\end{aligned}
		\label{F10}
	\end{equation}
	and similar identities for the pairs $k_2,k_2', \cdots k_p,k_p'$.
\end{remark}

\begin{remark}
	In the replica approach, one is usually interested in the  limit  where  the number of replicas $n \to 0$. It  is   however often easier to first think in terms of a  non-zero number of  of replicas and  to take the $n\to 0$ limit afterwards.  In this spirit, it is possible to generalize the above formulae (\ref{F7},\ref{F8}) following  a very similar calculation. 
	
	Defining the weighted overlaps  for non-zero numbers $n$ and $n'$ of replicas ($n$ and $n'$ are a priori arbitrary real numbers) by
	$$\langle Y_{k,k'}\rangle_{n,n'}  = { \Big\langle {\sum_{\nu }  y_\nu  \,  e^{
				-(\beta k + \beta' k')E_\nu}
			Z(\beta)^{n-k}  Z(\beta')^{n'-k'}} \Big\rangle  \over \langle Z(\beta)^n \ Z(\beta')^{n'} \rangle }$$

	It turns out that the expressions have somewhat simpler forms when the numbers $n$ and $n'$ take negative values  and one gets
	\begin{equation}
		\begin{aligned}
			\langle Y_{k,k'}\rangle_{n,n'} 
			=
			& {
				(-r)  \ \Gamma(-n) \ \Gamma(-n')
				\over \Gamma(k-n)\   \Gamma(k'-n')}
			\\ &  { \int_0^\infty dv  \ v^{k'-n' -1}
				\Psi\left( k + k' {\beta' \over \beta} ; v  \right)
				\ (-\psi(v))^{r-1}
				\over
				\int_0^\infty dv   \ v^{-n'-1}\
				(-\psi(v))^r }
		\end{aligned}
		\label{F9}
	\end{equation}
	where \begin{equation} 
		r={n \beta + n' \beta' \over \beta_c}   \ .
		\label{r-expression}
	\end{equation}
\end{remark}

\begin{remark}
	Using the following asymptotics of $\psi(v)$ which can be derived from expression (\ref{psi1})
	\begin{equation}
		\psi(v)  \simeq \left\{ 
		\begin{array}{lll}
			\Gamma\left(-{\beta_c \over \beta} \right) & \textrm{as} & v \to 0 \\
			\frac{\beta}{\beta'}\Gamma\left(-{\beta_c \over \beta'} \right)  v^{\beta_c \over \beta'} & \textrm{as}& v \to \infty
		\end{array}
		\right.
		\label{asymptotics}
	\end{equation}
	one can show that (\ref{F9}) reduces to (\ref{F7}) in the limit $n\to 0^-$ and $n' \to 0^-$.
\end{remark}

\section{Integer moments of the partition function at multiple temperatures}
	\label{sec:appendix-integer-moments}
The replica method  starts usually with  the calculation of  integer moments of the partition function.  In  a two or a multiple temperature case, these are  of  the form $ \left\langle Z(\beta_1)^{n_1} Z(\beta_2)^{n_2} Z(\beta_3)^{n_3} \cdots \right\rangle $ where $ n_1,n_2, n_3 \ldots $ are positive integers and $ \beta_1,\beta_2,\beta_3, \ldots $ are inverse temperatures. In this appendix we obtain the expressions (\ref{eq:Zn-sum-on-mu-exact}) and (\ref{eq:ZZnn-sum-on-mu-exact}) using  a generating function 
 defined for $ p $ temperatures as
\begin{equation} \label{eq:gf-multi-temp-dfnn}
 G(t_1,t_2,\ldots t_p)  = \left\langle \exp\left( -\sum_{i=1}^{p} t_i Z(\beta_i)\right) \right\rangle.
\end{equation}
For the REM  (see section \ref{sec:direct-calculation}) the partition function is given by
$$ Z(\beta) = \sum_{{\mathcal{C} }=1}^{2^N}    e^{-\beta E({\mathcal{C}})} $$
where the $2^N$ energies $E({\mathcal{C}})$ take   random values distributed according to $P(E)$ given in (\ref{eq:rem-disorder}). Then, because the $E({\mathcal{C}})$ are independent,
\begin{align*}
G(t_1,\cdots t_p )
& = \left[ \int P(E) d E \, \exp\left( - \sum_i t_i e^{-\beta_i E} \right) \right]^{2^N} 
 \\ & = \exp \left\{ 2^N \log \left(   \int P(E) d E \, \exp\Big( - \sum_i t_i e^{-\beta_i E} \Big)\right)  \right\} 
\end{align*}
which for large $N$ becomes
\begin{equation}
G(t_1,\cdots t_p )
\simeq \exp \left\lbrace 
\int_{-\infty}^{\infty} \rho(E) \left[ \exp \left(
-\sum_{i=1}^{p} t_i e^{-\beta_iE}
\right) -1 \right] dE 
\right\rbrace.
\label{Generating} 
\end{equation} 
(By this approximation, we in fact replace the REM by  a Poisson REM of density (see (\ref{eq:rem-disorder}))
\begin{equation}
\rho(E) = 2^N P(E)= 2^N  {1 \over \sqrt{N \pi J^2 }} \exp\left[ -{E^2 \over N J^2 }\right]
\ . 
\label{rho-def} 
\end{equation}
   Doing so the error is exponentially small in the system size $N$ as shown in the appendix of  \cite{Derrida_2015_Finite}).
The exponentials on the right hand side of (\ref{Generating}) can be expanded to obtain
\begin{multline} \label{eq:gf-multi-temp}
G(t_1,\cdots t_p )
= \sum_{r=0}^{\infty} \frac{1}{r!} \Biggl[ 
\mathop{\sum_{\mu_1=0}^{\infty} \cdots \sum_{\mu_p=0}^{\infty}}_{(\mu_1+ \cdots + \mu_p \geq 1)}
\frac{(-t_1)^{\mu_1}}{\mu_1!} \cdots \frac{(-t_p)^{\mu_p}}{\mu_p!} \\
\left< Z(\beta_1\mu_1+ \cdots + \beta_p \mu_p) \right>
\Biggr]^r
\end{multline}
where  we use the fact that for the REM (as well as for the Poisson REM) 
\begin{equation}
\left\langle Z(\beta) \right\rangle = \int_{-\infty}^{\infty} \rho(E) e^{-\beta E} \, dE.
\label{Zaverage}
\end{equation}
The general expression for integer moments at $ p $ temperatures is obtained by equating powers of $ t_i $ in the expansion of the right hand side of equation~\eqref{eq:gf-multi-temp-dfnn} with the right hand side of equation~\eqref{eq:gf-multi-temp}. Here  we give the three moments that are used in the main text.

The single temperature moments  are then  given by
\begin{multline} \label{eq:Zn-sum-on-mu-exact-appendix}
\left\langle Z(\beta)^n \right\rangle 
	= \sum_{r \geq 1} \frac{1}{r!} 
	\sum_{\{\mu_i \geq 1\}}
	C_{n,r} \left( \{\mu_i \}\right)
	\left\langle Z(\beta \mu_1) \right\rangle
		\left\langle Z(\beta \mu_2) \right\rangle
		\cdots
		\left\langle Z(\beta \mu_r) \right\rangle
\end{multline}
where we have defined
\begin{equation}
\sum_{\{\mu_i \geq 1\}}
	= \sum_{\mu_1 \geq 1} \sum_{\mu_2 \geq 1} 
	\cdots \sum_{\mu_r \geq 1} 
\end{equation}
and 
\begin{equation} \label{eq:C-dfnn-appendix}
	C_{n,r} \left( \{\mu_i \}\right) 
	= \frac{n!}{\mu_1! \mu_2! \cdots \mu_r!}
	\delta \left[ \sum_{i=1}^{r} \mu_i = n \right]
\end{equation}
The  Kronecker delta $ \delta \left[ \sum_{i=1}^{r} \mu_i = n \right] $ ensures that the $ \mu_i $ always sum to $ n $.

Similarly the two temperature moments  are  given by
\begin{multline} \label{eq:ZZnn-sum-on-mu-exact-appendix}
\left\langle Z(\beta)^{n} Z(\beta')^{n'} \right\rangle 
	= \sum_{r \geq 1} \frac{1}{r!} 
	\sum_{\{\mu_i \geq 0\}}
	\sum_{\{\mu'_i \geq 0\}}
	\theta \left[ \mu_i + \mu'_i \geq 1 \right] \, 
	C_{n,r} \left( \{\mu_i \}\right) \,  C_{n',r} \left( \{\mu'_i \}\right)\\
	\times
	\left\langle Z(\beta \mu_1 + \beta' \mu'_1) \right\rangle
	\left\langle Z(\beta \mu_2 + \beta' \mu'_2) \right\rangle
	\cdots
	\left\langle Z(\beta \mu_r + \beta' \mu'_r) \right\rangle
\end{multline}
where $ \theta \left[ \mu_i + \mu'_i \geq 1 \right] $ is unity if the inequality is satisfied for every $ i= 1,2, \ldots, r $ and zero otherwise.
The three temperature moments  are  given by
\begin{multline} \label{eq:ZZZnnn-sum-on-mu-exact-appendix}
\left\langle Z(\beta)^{n} Z(\beta')^{n'} Z(\beta'')^{n''} \right\rangle 
	= \sum_{r \geq 1} \frac{1}{r!} 
	\sum_{\{\mu_i \geq 0\}} \sum_{\{\mu'_i \geq 0\}} 
		\sum_{\{\mu''_i \geq 0\}}
	\theta \left[ \mu_i + \mu'_i + \mu''_i \geq 1 \right] \\
	\times
	C_{n,r} \left( \{\mu_i \}\right) C_{n',r}  \left( \{\mu'_i \}\right)
		C_{n'',r}  \left( \{\mu''_i \}\right)
 \ 	\left\langle Z(\beta \mu_1 + \beta' \mu'_1 + \beta'' \mu''_1) 
		\right\rangle\\
	\times
	\left\langle Z(\beta \mu_2 + \beta' \mu'_2 + \beta'' \mu''_2)
		\right\rangle
	\cdots
	\left\langle Z(\beta \mu_r + \beta' \mu'_r + \beta'' \mu''_r) \right\rangle
\end{multline}
\ \\ \ \\

As a special case of (\ref{eq:ZZnn-sum-on-mu-exact-appendix})
one has 
\begin{multline} 
\label{explanation1}
\left\langle Z(\beta)^{n-k} Z( k \beta) \right\rangle 
	= \sum_{r \geq 1} \frac{1}{r!} 
	\sum_{\{\mu_i \geq 0\}}
	\sum_{\{\mu'_i \geq 0\}}
	\theta \left[ \mu_i + \mu'_i \geq 1 \right] \, 
	C_{n-k,r} \left( \{\mu_i \}\right) \,  C_{1,r} \left( \{\mu'_i \}\right)\\
	\times
	\left\langle Z(\beta \mu_1 + k \beta \mu'_1) \right\rangle
	\left\langle Z(\beta \mu_2 + k \beta \mu'_2) \right\rangle
	\cdots
	\left\langle Z(\beta \mu_r + k \beta \mu'_r) \right\rangle
\end{multline}
In this case $n'=1$, therefore there is a single $\mu_i'=1$ all the others being $0$. Because of the symmetry between the indices $i$ in the previous formula, one can choose $\mu_1'=1$ and one gets
\begin{multline} \label{explanation2}
\left\langle Z(\beta)^{n-k} Z( k \beta) \right\rangle 
	= \sum_{r \geq 1} \frac{1}{(r-1)!} 
	\sum_{\{\mu_i \ge 1\}}
	C_{n,r} \left( \{\mu_i \}\right) \ 
		{(n-k) ! \over n!} {\mu_1 ! \over (\mu_1-k)!}
\\ 	\times
	\left\langle Z(\beta \mu_1 ) \right\rangle
	\left\langle Z(\beta \mu_2 ) \right\rangle
	\cdots
	\left\langle Z(\beta \mu_r ) \right\rangle
\end{multline}
where we take $ \frac{1}{(\mu_1-k)!} = 0 $ when $ \mu_1 <k $. 

\bibliographystyle{unsrtnatpjm}
\bibliography{pjm-master}

\end{document}